\documentclass[final,3p,times,twocolumn]{elsarticle}

\usepackage[labelfont=bf,justification=raggedright,singlelinecheck=false]{caption}
\usepackage{graphicx}
\usepackage{subcaption}

\usepackage{amssymb}

\usepackage{amsthm}
\usepackage{amsmath}
\usepackage{algorithm}
\usepackage[noend]{algpseudocode}

\usepackage{natbib}
\usepackage{lineno}

\usepackage{xcolor}

\newcounter{bla}

\journal{Computer Physics Communications}

\begin{document}
\begin{frontmatter}

\title{DynaPhoPy: A code for extracting phonon quasiparticles from molecular dynamics simulations}
\makeatletter
\author[a]{Abel Carreras\corref{author}}
\author[b]{Atsushi Togo}
\author[a,b]{Isao Tanaka}

\cortext[author] {Corresponding author.\\\textit{E-mail address:} abelcarreras83@gmail.com}
\address[a]{Department of Materials Science and Engineering, Kyoto University, Kyoto 606-8501, Japan}
\address[b]{Center for Elements Strategy Initiative for Structural Materials(ESISM), Kyoto University, Kyoto 606-8501, Japan}

\begin{abstract}
We have developed a computational code, \textsc{DynaPhoPy}, that allow us to
 extract the microscopic anharmonic phonon properties from molecular
 dynamics (MD) simulations using the normal-mode-decomposition technique
 as presented by Sun {\it et al.} [T. Sun, D. Zhang, R. Wentzcovitch, 2014]. Using this code
 we calculated the quasiparticle phonon frequencies and linewidths of
 crystalline silicon at different temperatures using both of first-principles and
 the Tersoff empirical potential approaches. In this work we show the dependence of these properties on
 the temperature using both approaches and compare them with reported experimental data 
 obtained by Raman spectroscopy [M. Balkanski, R. Wallis, E. Haro, 1983 and R. Tsu, J. G. Hernandez, 1982].  
\end{abstract}


\end{frontmatter}



{\bf PROGRAM SUMMARY}

\begin{small}
\noindent
{\em Manuscript Title:} 
DynaPhoPy: A code for extracting phonon quasiparticles from molecular dynamics simulations                                                   \\
{\em Authors:} Abel Carreras, Atsushi Togo and Isao Tanaka    \\
{\em Program Title:} DynaPhoPy                                \\
{\em Journal Reference:}                                      \\
{\em Catalogue identifier:}                                   \\
{\em Licensing provisions:} MIT License                       \\
{\em Programming language:} Python and C                      \\
{\em Computer:} PC and cluster computers                      \\
{\em Operating system:} UNIX/OSX                              \\
{\em RAM:} Depends strongly on number of input data (several Gb)    \\
{\em Number of processors used:} 1-16                         \\
{\em Supplementary material:}                                 
{\em Classification:} 7.8 Structure and Lattice Dynamics      \\
{\em External routines/libraries:} phonopy, numpy, matplotlib, 
scipy and h5py python modules. Optional: FFTW and Cuda  \\
{\em Subprograms used:}                                       \\
{\em Catalogue identifier of previous version:}*              \\
{\em Journal reference of previous version:}*                  \\
{\em Does the new version supersede the previous version?:}*   \\
{\em Nature of problem:}\\
 Increasing temperature, a crystal potential starts to deviate from the
 harmonic regime and anharmonicity is getting to be evident\cite{dove1993introduction}. 
 To treat anharmonicity, perturbation
 approach often describes successfully phenomena such as phonon lifetime and
 lattice thermal conductivity. However it fails when the system contains
 large atomic displacements.\\
{\em Solution method:}\\
Extracting the phonon quasiparticles from molecular dynamics (MD) simulations using the normal-mode-decomposition technique. 
   \\
{\em Reasons for the new version:}*\\
   \\
{\em Summary of revisions:}*\\
   \\
{\em Restrictions:}\\
Quantum effects of lattice dynamics are not considered.
   \\
{\em Unusual features:}\\
   \\
{\em Additional comments:}\\
   \\
{\em Running time:}\\
  It is highly dependent on the type of calculation requested.
  It depends mainly on the number of atoms in the primitive cell, the number of time steps of the MD simulation and the method employed to calculate the power spectra. Currently two methods are implemented in DyaPhoPy: The Fourier transform and the maximum entropy methods. The Fourier transform method scales to $\mathcal{O}$ [$N^2$] and the maximum entropy method scales to $\mathcal{O}$ [$N \times M$] where $N$ is the number of time steps and $M$ is the number of coefficients.
  
\end{small}

\section{Introduction}

Lattice dynamics calculations are a powerful tool to characterize the collective motion of atoms in crystalline materials. These calculations allow us to describe the thermodynamic behavior of crystals. 
For nearly harmonic crystals the harmonic approximation\cite{wallace1998thermodynamics} is commonly used. Under this approximation the temperature dependence of phonon frequencies is not described. To study this dependence it is necessary to consider the anharmonicity of crystals. 

When we expand the crystal potential into series with respect to the phonon coordinates, the weak anharmonicity is expected to be well described with lower order phonon-phonon interactions. These interactions induce phonon frequency shift and lifetime, which may be represented by the phonon quasiparticle picture. These phonon quasiparticles are observable experimentally, e.g, by Raman spectroscopy and neutron scattering.

In this work, we use molecular dynamics (MD) simulations to analyze the crystal anharmonicity as a function of temperature. In this analysis, the power spectrum of the mass-weighted velocity obtained from MD is fitted to model-spectral-function shapes to calculate both quasiparticle phonon frequencies and linewidths. For this purpose we developed a computational code in which the normal-mode-decomposition technique\cite{McGaughey2013} as reported by Sun \textit{et al.}\cite{Sun2014} is implemented. This technique allow us to extract the phonon quasiparticles from MD trajectories by projecting the atomic velocities onto the phonon eigenvectors.

We applied our code to silicon to obtain the quasiparticle phonon frequencies and linewidths at different temperatures. To run the MD simulations, we employed first-principles and the Tersoff empirical potential\cite{Tersoff1988} approaches. The first-principles calculations give us accurate energy landscapes, though they are computationally expensive. On the other hand, the use of empirical potentials allow us to run the MD simulations with longer simulation times and larger supercell sizes than using first-principles calculations, which is of major importance to densely sample the reciprocal space. 
As a result of our calculations, we found that the experimental phonon frequency shifts are well reproduced using both first-principles and the Tersoff empirical potential approaches for silicon. However, in the phonon linewidth calculations we observe a disagreement between both approaches. We shortly discuss these results by comparing them with the experiments by Raman spectroscopy available in literatures\cite{Balkanski1983, Tsu1982}.

\section{Anharmonic model}

In a perfect crystal, the atomic position ${\bf r}_{jl}$ can be written as
\begin{linenomath}
	\begin{equation} \label{eq:atomic_position}
		{\bf r}_{jl}(t) = {\bf r}^0_{jl} + {\bf u}_{jl}(t),
	\end{equation}
\end{linenomath}
where ${\bf r}^0_{jl}$ is the equilibrium position, and ${\bf u}_{jl}$ is the atomic displacement from ${\bf r}^0_{jl}$ at the lattice point $l$ and atomic index $j$ of each lattice point. Within the harmonic approximation the atomic displacement is described as a superposition of phonon normal modes:
\begin{linenomath}
	\begin{equation} \label{eq:harmonic}
		{\bf u}_{jl}(t) = \frac{1}{{\sqrt {{N}{m_j}} }}\sum\limits_{{\bf q}s} { {\bf e}_j({ \bf q}, s)\, {e^{i {\bf q} \cdot {\bf r}^0_{jl}}} u_{{\bf q}s}(t)},
	\end{equation}
\end{linenomath}
where $N$ is the number of lattice points, $m_j$ is the atomic mass, ${\bf e}$ is the phonon eigenvector, $\bf q$ and $s$ are the wave vector and branch, respectively, and $u_{{\bf q}s}(t)$ is the displacement in the phonon coordinate. These phonon modes are obtained as the solution of the eigenvalue problem of the dynamical matrix 
\begin{linenomath}
	\begin{equation} \label{eq:eigenvalue_equation}
		{\bf D}({\bf q})\, {\bf e}({\bf q},s)= \omega_{{\bf q}s}^2\, {\bf e}({\bf q},s),
	\end{equation}
\end{linenomath}
 where $\omega_{{\bf q}s}$ is the phonon frequency and ${\bf D}({\bf q})$ is the dynamical matrix, whose elements are defined as 
\begin{linenomath}
	\begin{equation} \label{eq:dynamical_matrix}
		{D}_{\alpha \beta } (jj',{\bf q } ) = \frac{1}{(m_j m_{j'} )^{1/2}}\sum\limits_{l'} {\Phi_{\alpha \beta}\left( {\begin{array}{*{20}{c}}
			{jj'}\\
			{0l'}
			\end{array}} \right) e^{i{\bf q } ( {\bf r}_{j'l'}^0 - {\bf r}^0_{j0} )}},
	\end{equation}
\end{linenomath}
where $\alpha$ and $\beta$ are the indices of the Cartesian coordinates. $\Phi$ is the harmonic force constants matrix whose elements are given by
\begin{linenomath}
	\begin{equation} \label{eq:force_constants}
		{\Phi_{\alpha \beta}\left( {\begin{array}{*{20}{c}}
			{jj'}\\
			{ll'}
			\end{array}} \right) = \frac{\partial^2 W}{\partial u_{j l \alpha }  \partial u_{ j' l' \beta}}}.
	\end{equation}
\end{linenomath}
where $W$ is the crystal potential energy. 

To include the anharmoncity we use the phonon quasiparticle model\cite{Zhang2014}. In this model, the atomic velocity ${\bf v}_{jl}(t)$ is defined to be similar to Eq. (\ref{eq:harmonic}) as
\begin{linenomath}
	\begin{equation} \label{eq:harmonic_velocity}
		{\bf v}_{jl}(t) = \frac{1}{{\sqrt {{N}{m_j}} }}\sum\limits_{{\bf q}s}  { {\bf e}_j({ \bf q}, s)\, {e^{i {\bf q} \cdot {\bf r}_{jl}}} v_{{\bf q}s}(t)},
	\end{equation}
\end{linenomath}
where $v_{{\bf q}s}(t)$ is the velocity of the phonon quasiparticle. We consider that the phonon quasiparticle is a good model if the power spectrum of $v_{{\bf q}s}(t)$, i.e., the Fourier transform of the autocorrelation function of $v_{{\bf q}s}(t)$, has a well defined spectral shape.
This power spectrum is the central information that we want to compute and it is given as

\begin{linenomath}
	\begin{equation} \label{eq:powerspectrum}
	   G_{{\bf q}s}(\omega) = 2 \int_{-\infty}^\infty  {\left\langle {{v_{{\bf q}s}^*}\left( 0 \right) v_{{\bf q}s} \left( \tau \right)} \right\rangle {e^{i \omega \tau}}d\tau}, 
	\end{equation}
\end{linenomath}
where $G_{{\bf q}s}(\omega)$ is the one-sided power spectrum and $\langle {{v_{{\bf q}s}^*}\left( 0 \right) v_{{\bf q}s} \left( \tau \right)}\rangle$ is the autocorrelation function of $v_{{\bf q}s}(t)$ defined as
\begin{linenomath}
	\begin{equation} \label{eq:autocorrelation}
	   \langle {{v_{{\bf q}s}^*}\left( 0 \right) v_{{\bf q}s} \left( \tau \right)}\rangle = \lim_{t'\rightarrow\infty} \frac{1}{t'}\int_0^{t'}{v_{{\bf q}s}(t+\tau)\, v_{{\bf q} s}^*(t)} dt.
	\end{equation}
\end{linenomath}
\section{Methodology}

We use the normal-mode-decomposition technique to obtain $v_{{\bf q}s}(t)$ from MD simulations. In this technique, the atomic velocities are projected onto a wave vector $\bf q$ as
\begin{linenomath}
  \begin{align} \label{eq:wave_vector_proj}
    {\bf v}_{j}^{\bf q}(t) = \sqrt{\frac{m_j}N}\sum\limits_l {{e^{ - i{\bf{q}} \cdot {{\bf{r}}^0_{j l}}}}{\bf v}_{j l}(t)},
  \end{align}
\end{linenomath}
Then, ${\bf v}_{j}^{\bf q}(t)$ are projected onto a phonon eigenvector ${\bf e}({\bf q},s)$ by 
\arraycolsep=5pt\def\arraystretch{1.5}
\begin{linenomath}
  \begin{align} \label{eq:eigenvector_proj}
	v_{{\bf q}s}(t) = \sum_j {{\bf v}_j^{\bf q}(t) \cdot {\bf e}^*_j ({\bf q},s)}. 
  \end{align}
\end{linenomath}
Finally, the power spectrum of $v_{{\bf q}s}(t)$ is calculated using Eq. (\ref{eq:powerspectrum}).
Sun and Allen showed the approximate expression of this power spectrum to have a Lorentzian function shape\cite{Sun2010a}. This is expected to work well when the phonon frequency shift and linewidth are both small.
Based on this approximation, we employ a Lorentzian function form to fit $G_{{\bf q}s}(\omega)$: 
\begin{linenomath}
	\begin{equation} \label{eq:lorentzian_fit}
		G_{{\bf q}s}(\omega) \approx \frac{\langle| v_{{\bf q}s}(t)|^2\rangle  }{{ \frac{1}{2} \gamma_{{\bf q}s} \pi \left({1 + {{\left( {\frac{{\omega - {\tilde \omega_{{\bf q}s}}}}{\frac{1}{2}\gamma_{{\bf q}s}}} \right)}^2}} \right)}}.
 	\end{equation}
\end{linenomath}
From this fitting the quasiparticle phonon frequency $\tilde \omega_{{\bf q}s}$ is determined as the peak position and the phonon linewidth $\gamma_{{\bf q}s}$ as the full width at half maximum (FWHM).

The wave-vector-projected power spectrum $G_{\bf q}(\omega)$ and full power spectrum $G(\omega)$ are written as

\begin{linenomath}	
	\begin{equation} \label{eq:wave_dos}
	   G_{\bf q}(\omega) = 2 \sum\limits_{j \alpha}  {\int_{-\infty}^\infty  {\left\langle {{v_{j \alpha}^{\bf q*}}\left( 0 \right) v_{j \alpha}^{\bf q}\left( \tau \right)} \right\rangle {e^{i \omega \tau}}d\tau}}
	\end{equation}
\end{linenomath}
and
\begin{linenomath}	
	\begin{equation} \label{eq:full_dos}
	   G(\omega) = 2 \sum\limits_{j l \alpha}  {\int_{-\infty}^\infty  {\left\langle {{\tilde v_{j l \alpha}^*}\left( 0 \right) \tilde v_{j l \alpha}\left( \tau \right)} \right\rangle {e^{i \omega \tau}}d\tau}},
	\end{equation}
\end{linenomath}
respectively. Since the phonon eigenvectors ${\bf e}({\bf q},s)$ are orthonormal, Eqs. (\ref{eq:powerspectrum}), (\ref{eq:wave_vector_proj}) and (\ref{eq:full_dos}) satisfy the following relation:

\begin{linenomath}	
	\begin{equation} \label{eq:ps_decomposition}
	   G(\omega)=\sum\limits_{\bf q} { G_{\bf q}(\omega) } = \sum\limits_{{\bf q} s}  { G_{{\bf q}s}(\omega) }.
	\end{equation}
\end{linenomath}
This relation allow us to define the total power of $G(\omega)$ as
\begin{linenomath}
	\begin{equation} \label{eq:lorentzian_pos}
		\int_0^\infty {G(\omega)} d\omega = \sum_{{\bf q}s} \int_0^\infty {G_{{\bf q}s}(\omega)} d\omega =\sum_{{\bf q}s} \langle|v_{{\bf q}s}(0)|^2\rangle,
 	\end{equation}
\end{linenomath}
where
\begin{linenomath}
	\begin{equation} \label{eq:lorentzian_pos}
		\langle|v_{{\bf q}s}(0)|^2\rangle = \lim_{t'\rightarrow\infty} \frac{1}{t'}\int_0^{t'}{v_{{\bf q}s}(t)v_{{\bf q} s}^*(t)}dt.
 	\end{equation}
\end{linenomath}
This shows that the total power of $G(\omega)$ is related with the kinetic energy as  
\begin{linenomath}	
	\begin{equation} \label{eq:total_power_int}
	   \int_0^\infty {G(\omega) d\omega} = 2 \langle K \rangle,
	\end{equation}
\end{linenomath}
since
\begin{linenomath}
	\begin{equation} \label{eq:vel_p_coordinate}
		\sum_{{\bf q}s} { \langle|v_{{\bf q}s}(0)|^2\rangle} =  \sum_{jl}{ m_j\langle|{\bf v}_{jl}(0)|^2\rangle} = 2 \langle K \rangle,
 	\end{equation}
\end{linenomath}
where ${\bf v}_{jl}(t)$ is the atomic velocity and $\langle K \rangle$ is the average vibrational kinetic energy of the system that we are interested in. At the classical limit,

\begin{linenomath}
	\begin{equation} \label{eq:kinetic_classical_limit}
		\langle K \rangle = \frac{3}2 Nn_ak_BT
 	\end{equation}
\end{linenomath}
where $n_a$ is the number of atoms per lattice point and $k_B$ the Boltzmann constant.

Within the harmonic approximation, the phonon density of states (DOS) is related to the Fourier transform of the velocity autocorrelation function within a constant factor\cite{dove1993introduction, Lee1993}. We employ this relation in the phonon quasiparticles model along with the result in Eq. (\ref{eq:ps_decomposition}) to define the quasiparticle DOS $g(\omega)$ as 
\begin{linenomath}
	\begin{equation} \label{eq:normalized_dos}
	   g(\omega) = \frac{1}{3Nn_a k_BT}G(\omega).
	\end{equation}
\end{linenomath}

\section{Software overview}

\textsc{DynaPhoPy} is mainly written in Python and its performance bottle-neck is treated by C. The purpose of this software is to extract quasiparticle phonon frequencies and linewidths from MD trajectories using the normal-mode-decomposition technique\cite{McGaughey2013}. At the present time \textsc{DynaPhoPy} prepares interfaces to \textsc{VASP}\cite{vasp1996,vasp1999} and \textsc{LAMMPS}\cite{Plimpton1995a} MD trajectories and relies on \textsc{Phonopy}\cite{phonopy} to calculate the phonon eigenvectors. To do this calculation it requires the information about the crystal structure and the forces on the atoms.
To calculate the quasiparticle phonon frequencies and linewidths, first, the mass-weighted velocity is projected onto a wave vector $\bf q$ to obtain ${\bf v}^{\bf q}(t)$ (Eq. (\ref{eq:wave_vector_proj})). Next, \textsc{DynaPhoPy} interfaces with \textsc{Phonopy} to obtain the phonon eigenvector at the wave vector $\bf q$ and the band index $s$, and ${\bf v}^{\bf q}(t)$ is projected onto this phonon eigenvector to obtain ${\bf v}_{{\bf q}s}(t)$ (Eq. (\ref{eq:eigenvector_proj})).
Finally, the power spectrum of ${\bf v}_{{\bf q}s}(t)$ is calculated (Eq. (\ref{eq:powerspectrum})) and fitted to the Lorentzian function (Eq. (\ref{eq:lorentzian_fit})). To increase the numerical accuracy of the power spectrum \textsc{DynaPhoPy} averages the power spectra at symmetrically equivalent wave vectors. 

In addition, \textsc{DynaPhoPy} uses the quasiparticle phonon frequencies $\tilde \omega_{{\bf q}s}$ calculated at commensurate $\bf q$-points corresponding to the chosen supercell size to interpolate $\tilde \omega_{{\bf q}s}$ at incommensurate $\bf q$-points. To perform this we define a crudely renormalized force constants matrix $\tilde\Phi$ whose elements are
\begin{linenomath}
	\begin{equation}\label{eq:renormalized_force_constants}
		\tilde \Phi_{\alpha \beta}\left( {\begin{array}{*{20}{c}}
			{jj'}\\
			{0l'}
			\end{array}} \right) =\frac{\sqrt{m_j m_{j'}}}N \sum_{\bf q}{\tilde D_{\alpha \beta}(jj',{\bf q})\, e^{-i{\bf q } ( {\bf r}^0_{jl} - {\bf r}^0_{j0} )}},
	\end{equation}
\end{linenomath}
with
\begin{linenomath}
	\begin{equation}\label{eq:renormalized_dynamical_matrix}
		\tilde D_{\alpha \beta}(jj',{\bf q})= \sum_s{\tilde \omega^2_{{\bf q}s}e_{j\alpha}({\bf q},s)e^*_{j'\beta}({\bf q},s)}.
	\end{equation}
\end{linenomath}
The renormalized force constants are useful to compute the thermodynamic properties including the anharmonic correction\cite{Allen2015}. In this work we do not use them since we want to focus on the analysis of the frequency shifts and linewidths, which is presented in the following sections.

\section{Computational details} \label{sec:test}

Crystalline silicon is a widely studied material, composed by a single element and packed in a cubic structure at ambient condition. Due to the high importance of its properties and applications there is a large amount of information available in literatures, both theoretical and experimental. This makes silicon a suitable candidate for benchmarking our code.

\begin{table}[h]
  \caption{Structural and computational parameters used in the first-principles MD simulations of silicon with \textsc{VASP} code.}
  \centering
 \begin{tabular}{p{35mm}p{35mm}}
    \hline \hline
    Parameters &  \\
    \hline
    Lattice parameter & 5.45 \AA \\
    k-point sampling & Monkhorst-Pack\cite{Monkhorst1976}  2x2x2 \\
    Methodology & Density functional \newline theory and GGA (PBESol\cite{PBESol2008})  \\
    Thermostat & Nos\'e-Hoover\cite{Hoover1695} \\
    Cutoff energy & 300 eV \\
    Number of atoms & 64 \\
    Number of time  steps & 500000  \\
    Relaxation time steps & 20000 \\
    Time step length & 2.0 fs \\
    \hline \hline
  \end{tabular}
  \label{table:silicon_params_vasp}
\end{table}

\begin{table}[h]
  \caption{Structural and computational parameters employed in the MD simulations of silicon using the Tersoff empirical potential with \textsc{LAMMPS} code.}
  \centering
 \begin{tabular}{p{35mm}p{35mm}}
    \hline \hline
    Parameters &  \\
    \hline
    Lattice parameter & 5.45 \AA \\
    Empirical potential & Tersoff\cite{Tersoff1988} \\
    Thermostat & Nos\'e-Hoover\cite{Hoover1695} \\
    Number of atoms & 64 and 512 \\
    Number of time  steps & 1000000  \\
    Relaxation time steps & 500000 \\
    Time step length & 1.0 fs \\
    \hline \hline
  \end{tabular}
  \label{table:silicon_params_lammps}  
\end{table}

In this work we present the quasiparticle phonon frequencies and linewidths of silicon as a function of temperature. For this purpose we used MD simulations computed using two different approaches, i.e, first-principles and the Tersoff empirical potential, for which \textsc{VASP}\cite{vasp1996,vasp1999} and \textsc{LAMMPS}\cite{Plimpton1995a} codes were employed, respectively. In Tables \ref{table:silicon_params_vasp} and \ref{table:silicon_params_lammps} we summarize the major parameters used to perform these calculations.

First-principles calculations were used as references with respect to calculations by the Tersoff empirical potential, since we expect them to have the higher level of prediction. MD simulations using the empirical potential are computationally less demanding, therefore we use them to analyze the dependency of the MD supercell size on the quasiparticle phonon properties.
\textsc{Phonopy} code was used to obtain the phonon eigenvectors. Harmonic force constants were calculated by the finite displacement method using the 2x2x2 supercell of the conventional unit cell. We also employed \textsc{Phonopy-qha} to calculate the thermal expansion contribution to the frequency shift based on the quasi-harmonic approximation (QHA). 

To calculate the power spectra we used two different methods: the Fourier transform (FT) method (\ref{sec:dft}) and the maximum entropy (ME) method (\ref{sec:mem}). According to our tests, the power spectra obtained using the ME method gives a smoother profile, however its calculation requires a larger number of MD time steps than that of the FT method. In small supercells we observed that the frequency shifts and linewidths obtained from $G_{{\bf q}s}(\omega)$ calculated using the MEM method are slightly overestimated with respect to those calculated using the FT method. In contrast, for the calculation of the total power of $G(\omega)$ we obtain similar results using both methods.
In this study we used the ME method to calculate the full power spectrum $G(\omega)$ and the FT method to calculate the frequency shift and linewidth from $G_{{\bf q}s}(t)$.

\section{Results and discussion}\label{sec:test2}

\begin{figure}[h]
  \begin{center}
    \begin{subfigure}[h]{0.5\textwidth}
      \caption{}
      \includegraphics[width=\linewidth]{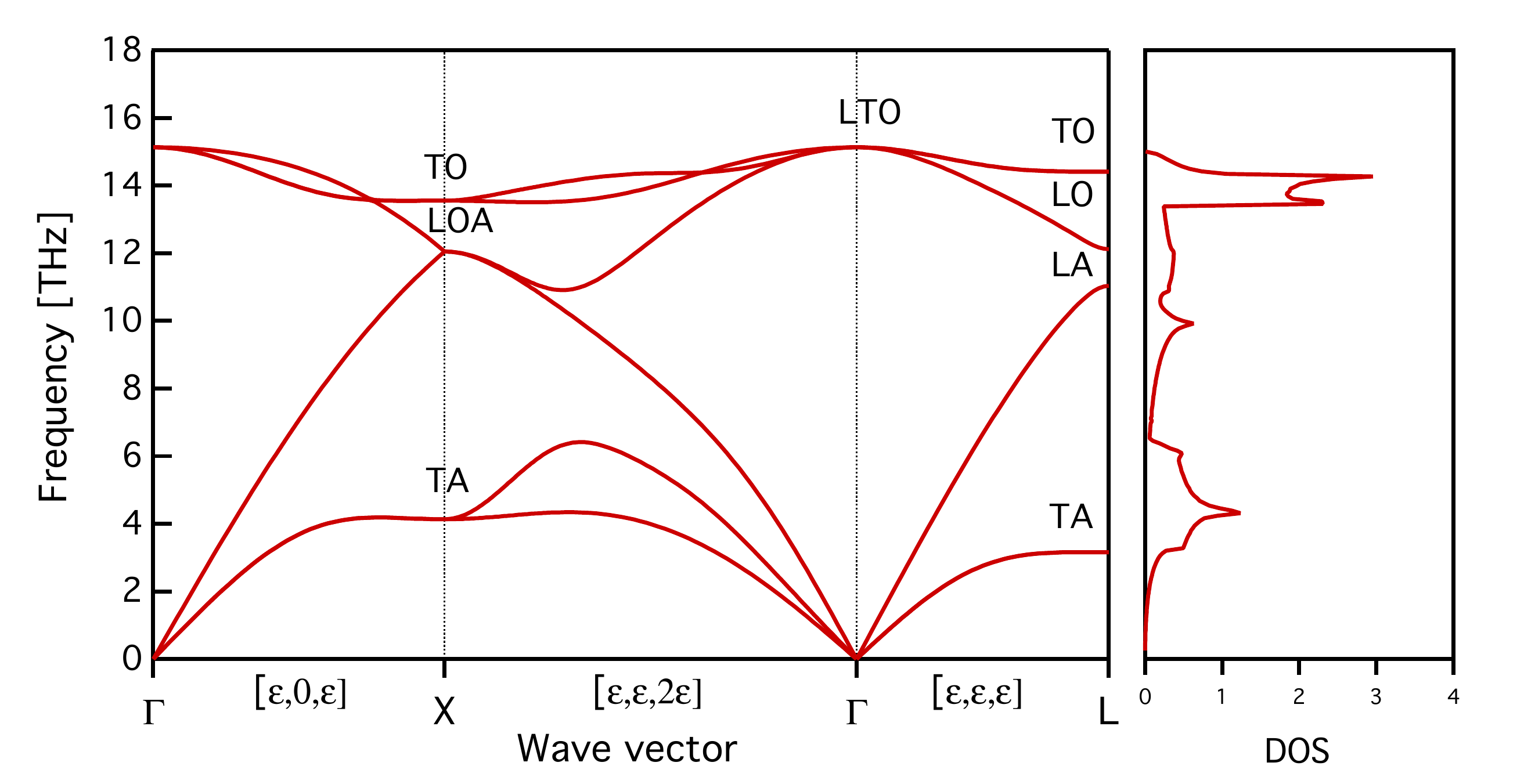}
    \end{subfigure}
    \begin{subfigure}[h]{0.5\textwidth}
      \caption{}
      \includegraphics[width=\linewidth]{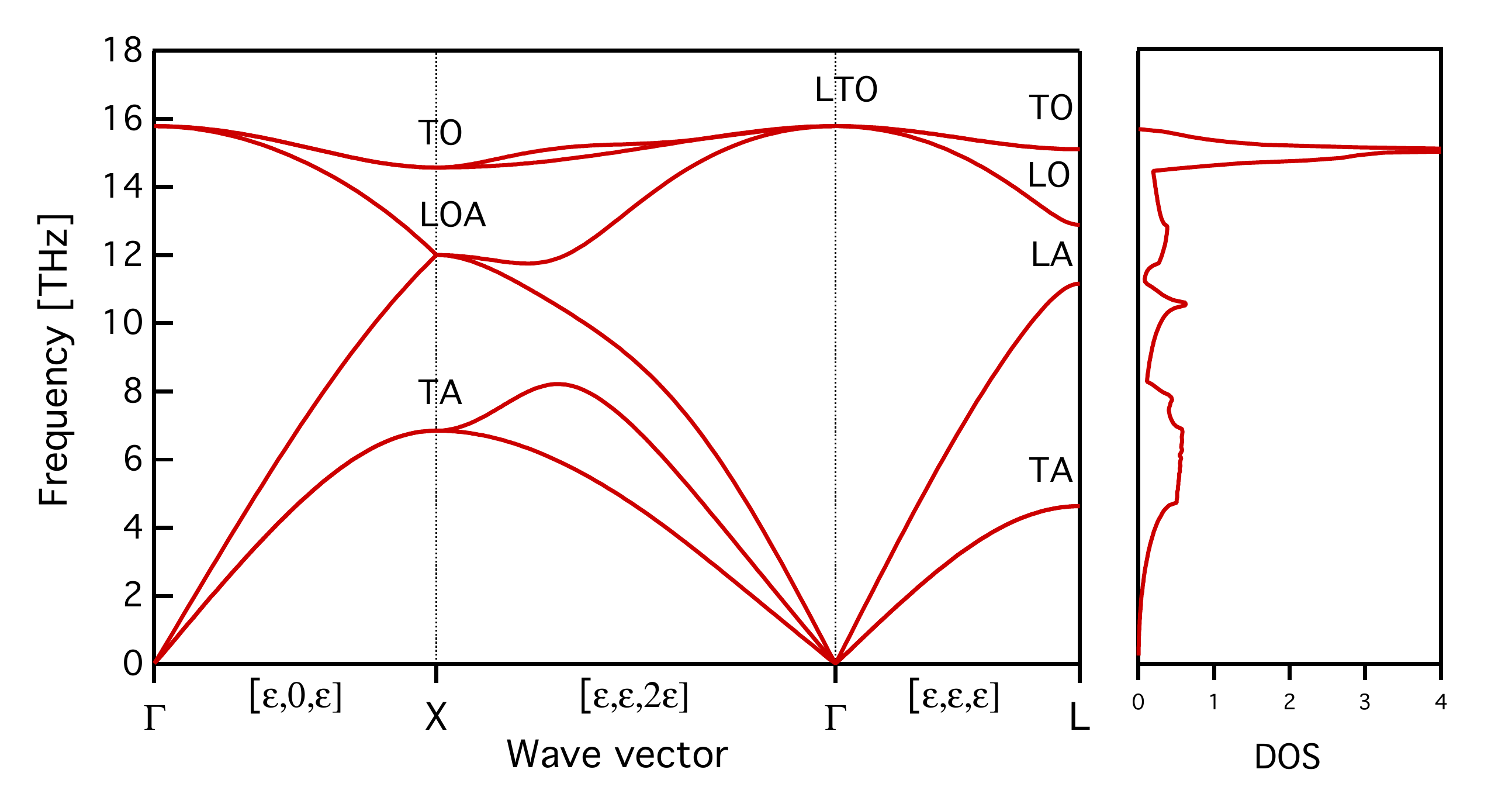}
    \end{subfigure}
    \caption{Harmonic phonon band structure (left) and DOS (right) of silicon calculated using (a) first-principles and (b) the Tersoff empirical potential approaches. The labels transversal ($T$),  longitudinal ($L$), acoustic ($A$) and optical ($O$) indicate the phonon branches. Phonon DOS is normalized to the number of phonon modes in the primitive cell.}
    \label{fig:si_disp_phon}
   \end{center}
\end{figure}

Fig. \ref{fig:si_disp_phon} shows the phonon band structures and DOS obtained using harmonic lattice dynamics calculations, where no temperature effect is considered. We use them as simple reference points of the band structure shapes to discuss the following results. 

In Sec. \ref{sec:power_spectra} we show the power spectra $G(\omega)$, $G_{\bf q}(\omega)$ and $G_{{\bf q}s}(\omega)$ obtained from the MD simulations. From $G(\omega)$ we check the convergence of the MD simulations by calculating its total power, which is related to the average kinetic energy (Eq. (\ref{eq:total_power_int})), and comparing it with the classical limit. In $G_{\bf q}(\omega)$ we observe the different peaks that correspond to quasiparticle phonon modes with wave vector ${\bf q}$. We assign these peaks in $G_{\bf q}(\omega)$ to phonon modes at commensurate $\bf q$-points in Fig. \ref{fig:si_disp_phon}. Analyzing the power spectra $G_{{\bf q}s}(\omega)$ we observe that the quasiparticle phonon modes are well separated using the phonon eigenvectors basis.

In Sec. \ref{sec:frequency_shifts} we show the frequency shifts obtained from MD simulations as a function of temperature. We include the contribution of the thermal expansion to the frequency shift employing the QHA. Then, we compare the results obtained using first-principles and the Tersoff empirical potential approaches with the Raman experiments, and finally we analyze the effect of the supercell size on the computed results. 
 
In Sec. \ref{sec:linewidths} we show the phonon linewidths obtained from MD simulations using the two different approaches and analyze their dependence on the supercell size.

\subsection{Power spectra} \label{sec:power_spectra} 
\begin{figure}[h]
  \begin{center}
    \begin{subfigure}[h]{0.5\textwidth}
      \caption{}
      \includegraphics[width=\linewidth]{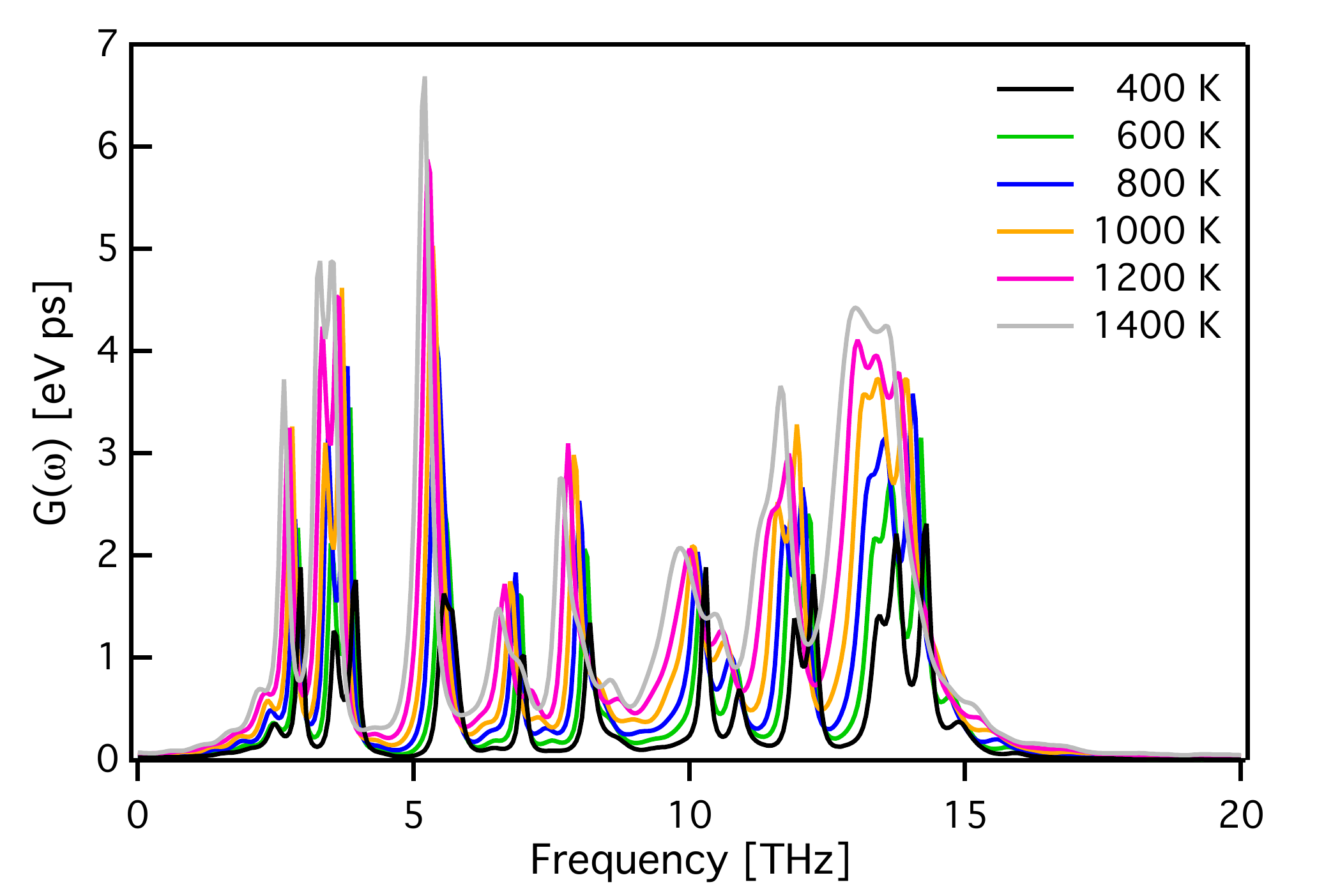}
    \end{subfigure}
    \begin{subfigure}[h]{0.5\textwidth}
      \caption{}
      \includegraphics[width=\linewidth]{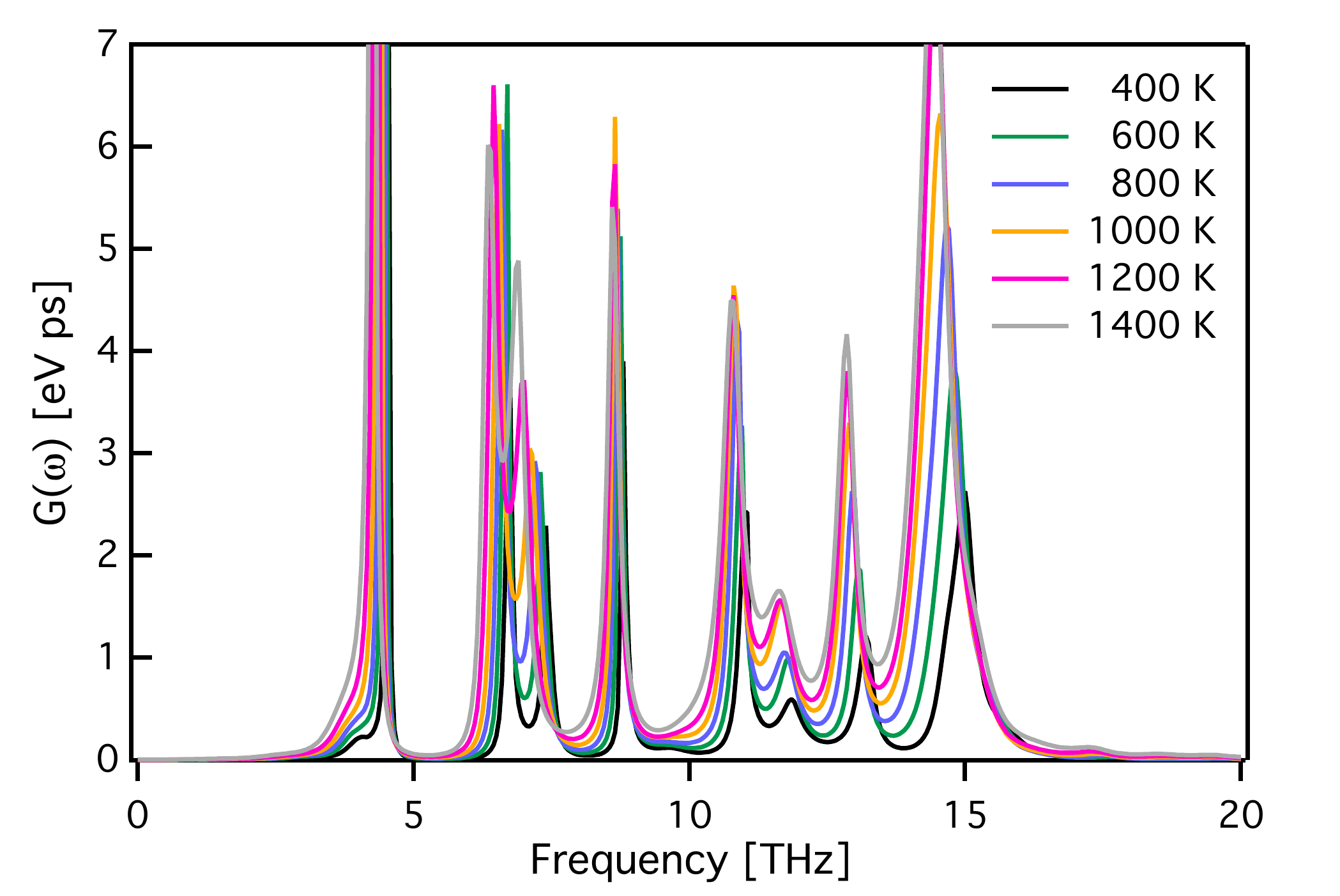}
    \end{subfigure}
    \caption{Full power spectra $G(\omega)$ of silicon obtained from MD simulations calculated using (a) first-principles and (b) the Tersoff empirical potential approaches. The power spectra were calculated using the ME method. The supercell of 64 atoms was used.}
    \label{fig:si_full_powerspectra}
   \end{center}
\end{figure}

The full power spectra $G(\omega)$ at different temperatures calculated using first-principles and the Tersoff empirical potential approaches are shown in Fig. \ref{fig:si_full_powerspectra}. These power spectra correspond to the superposition of the peaks of quasiparticle phonon modes at commensurate $\bf q$-points (Eq. (\ref{eq:ps_decomposition})) of the supercell of 64 atoms. The area underneath correspond to the total vibrational kinetic energy while the position of each peak indicates the frequency of a phonon quasiparticle. However, in this power spectrum the peaks are highly overlapped and the individual analysis of each phonon quasiparticle cannot be performed.
As the temperature increases the peaks become wider and their position is displaced to lower frequencies. This behavior is due to the anharmonicity.
The number of peaks is small since the number of phonon modes represented in each MD simulation box is proportional to the number of atoms in the supercell. Therefore, by our choices of supercell sizes, the shapes of $G(\omega)$ largely differ from the DOS shown in Fig. \ref{fig:si_disp_phon}. The differences observed between the results obtained using first-principles and the Tersoff empirical potential are relatively small but they are more noticeable in $G(\omega)$ than in the phonon band structure. This happens because $G(\omega)$ only includes information of phonon at commensurate $\bf q$-points while the phonon band structure and DOS shown in Fig. \ref{fig:si_disp_phon} include phonon interpolated at non-commensurate $\bf q$-points.

\begin{figure}[h]
  \begin{center}
    \includegraphics[width=1\linewidth]{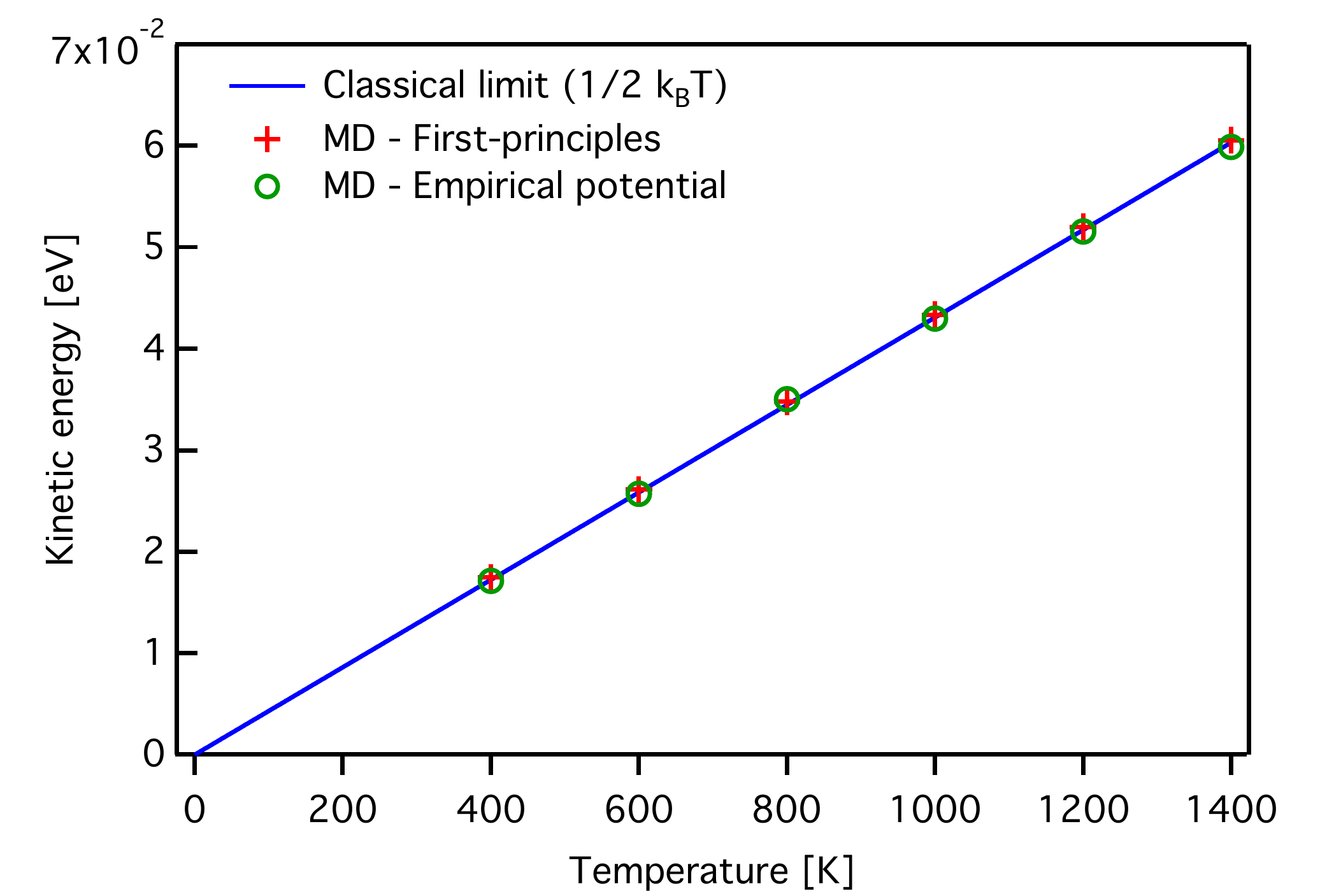}
        \caption{Kinetic energies per phonon mode of silicon obtained from MD simulations as a function of temperature compared with the classical limit ($\frac{1}2k_BT$). MD simulations were run by using first-principles (red crosses) and the Tersoff empirical potential (green circles) approaches with the supercell of 64 atoms. The ME method was employed to calculate the full power spectra $G(\omega)$ kinetic energies. The blue line depicts the classical limit.}
        \label{fig:si_full_boltzmann}  
   \end{center}
\end{figure}

 In Fig. \ref{fig:si_full_boltzmann} we compare the average kinetic energies per phonon mode obtained from $G(\omega)$ as
\begin{linenomath}	
	\begin{equation}\label{eq:integration_kinetic}
		\frac{1}{2N_l}\int_0^\infty{G(\omega)d\omega}
	\end{equation}
\end{linenomath}
with the classical limit, where $N_l$ is the number of degrees of freedom of the system. In \textsc{LAMMPS} calculations we employed $N_l=3n_a$ while in \textsc{VASP} we used $N_l=3n_a-3$ due to the different treatments of rigid translation in these codes. In contrast to LAMMPS, \textsc{VASP} adds three additional restrictions to the system by keeping the center of mass constant to avoid the lattice drift during the MD simulation. The results obtained for the average kinetic energy show an excellent agreement with the classical limit at all temperatures in both VASP and LAMMPS calculations. This indicates that the full power spectrum is well reproduced using the ME method.

\begin{figure}[h]
  \begin{center}
    \includegraphics[width=1\linewidth]{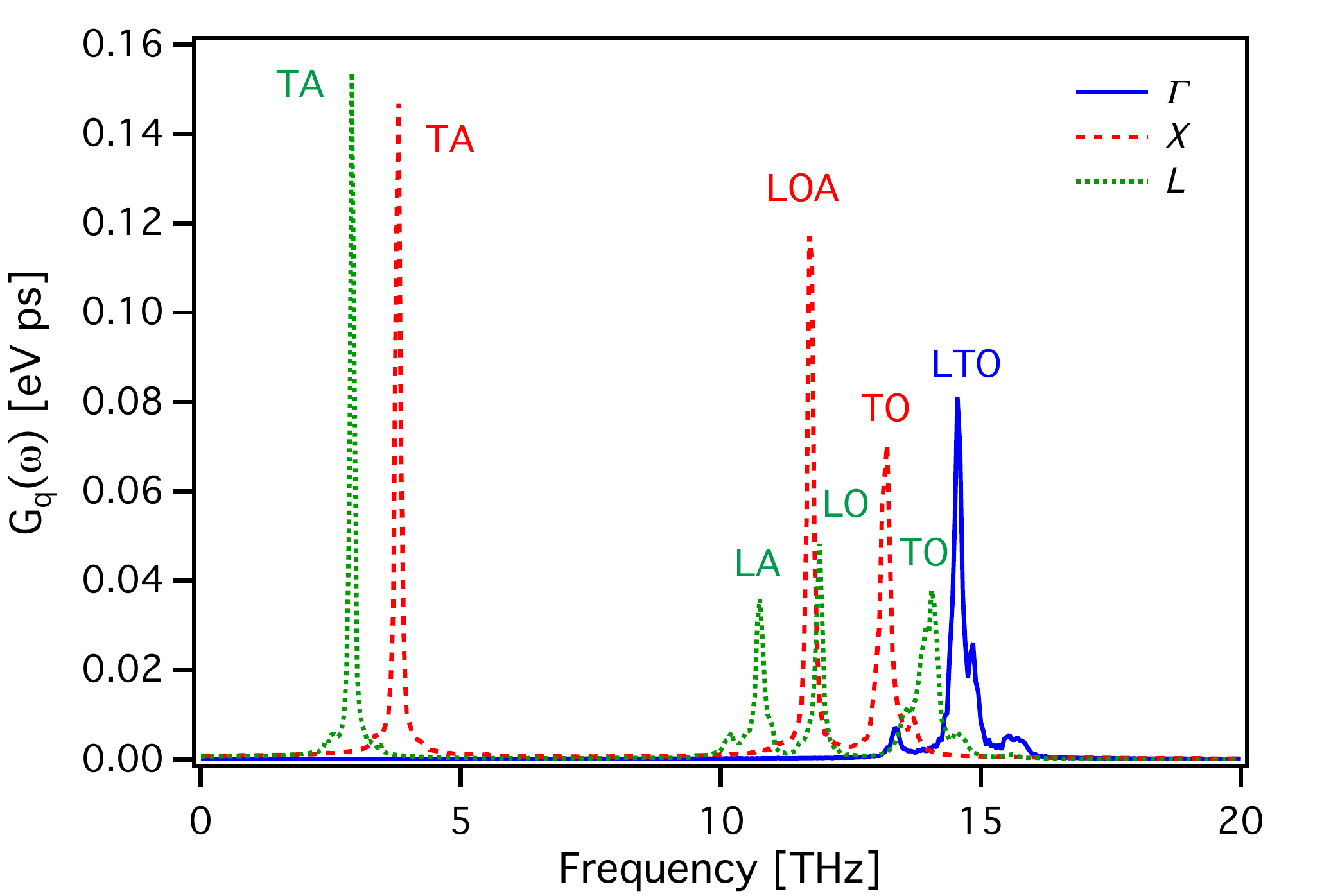}
        \caption{Wave-vector-projected power spectra $G_{\bf q}(\omega)$ at $\Gamma$ (blue), $X$ (red) and $L$ (green) points. The power spectra are calculated from first-principles MD simulations at 800 K using the FT method. The labels transversal ($T$),  longitudinal ($L$), acoustic ($A$) and optical ($O$) indicate the corresponding phonon branches.}
        \label{fig:wave_powerspectra}
   \end{center}
\end{figure}

In Fig. \ref{fig:wave_powerspectra} we show the wave-vector-projected power spectra $G_{\bf q}(\omega)$ of silicon at $\Gamma$, $X$, and $L$ points obtained from the first-principles MD simulations at 800 K using the FT method. In these power spectra we can easily distinguish the peaks corresponding to each individual quasiparticle phonon mode. From the position of each peak we extract its phonon frequency.  We use this information to assign each peak to a branch of the phonon band structure shown in Fig 1. 

At $\Gamma$ point (solid blue curve) we observe only one peak that corresponds to the LTO branch. The acoustic mode at $\Gamma$ point should not be observed as shown in Fig. \ref{fig:wave_powerspectra}.  This acoustic phonon mode at $\Gamma$ point corresponds to lattice translations which are restricted in \textsc{VASP} MD simulations.
At $X$ (dashed red curve) and $L$ (dotted green curve) points we observe three and four peaks respectively. This agrees with the number of phonon branches obtained in the phonon band structure in Fig. \ref{fig:si_disp_phon}. 

\begin{figure*}[h]
        \centering
        \begin{subfigure}[h]{0.35\textwidth}
                \caption{First-principles}
                \includegraphics[width=\textwidth]{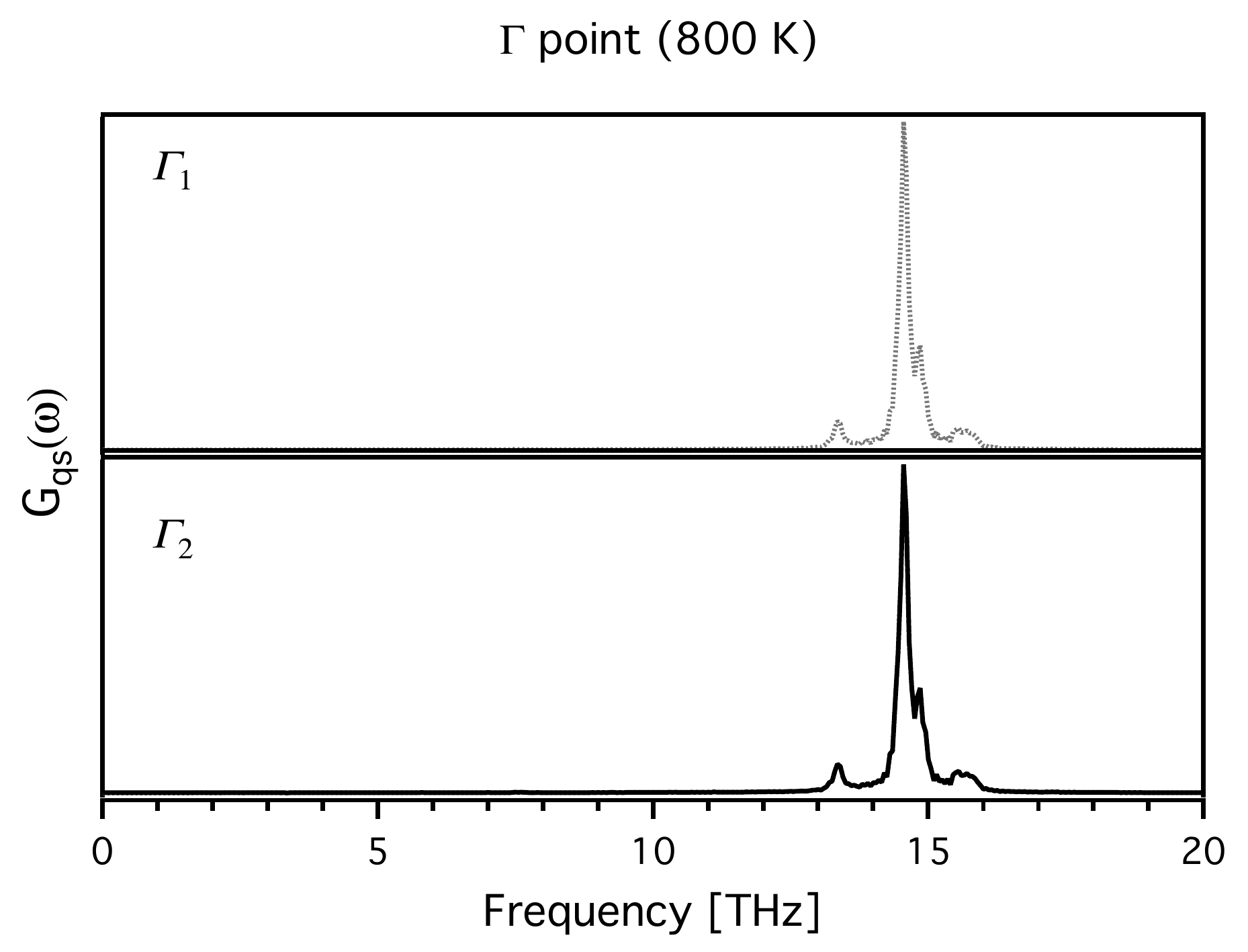}
        \end{subfigure}
        \hspace{2cm}
        \begin{subfigure}[h]{0.35\textwidth}
                \caption{Empirical potential}
                \includegraphics[width=\textwidth]{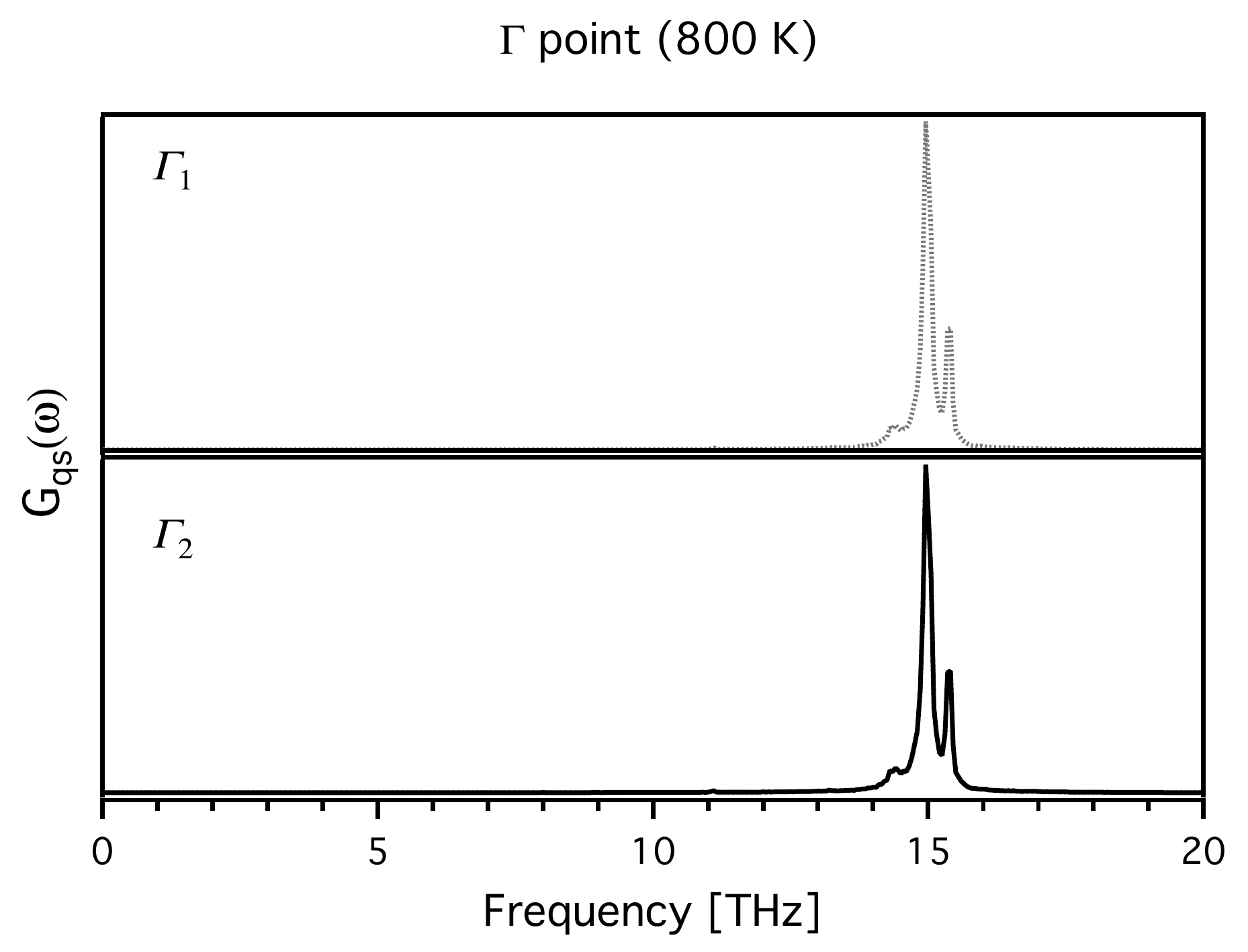}
        \end{subfigure}
        \begin{subfigure}[h]{0.35\textwidth}
                \includegraphics[width=\textwidth]{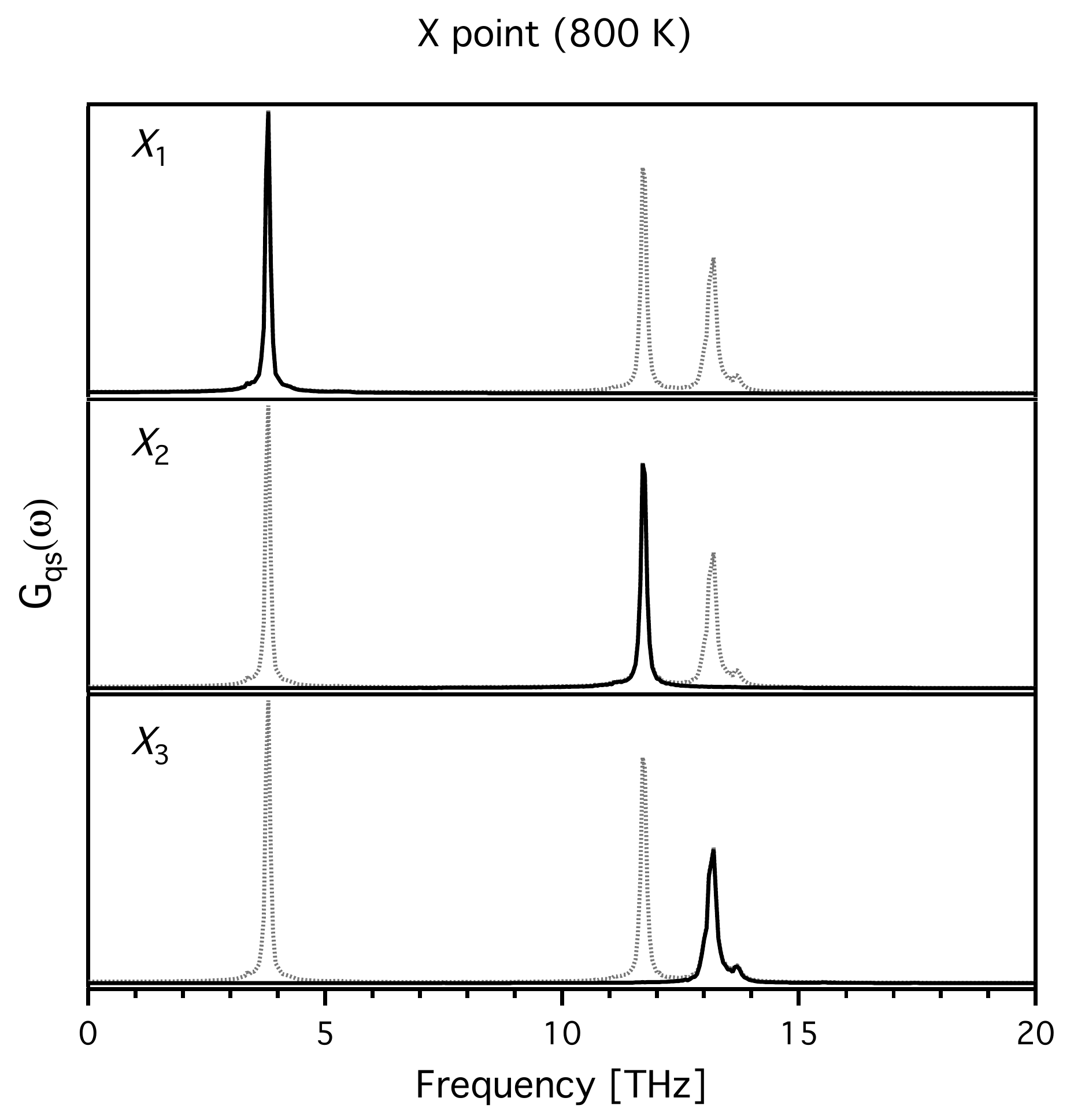}
        \end{subfigure}
        \hspace{2cm}
        \begin{subfigure}[h]{0.35\textwidth}
                \includegraphics[width=\textwidth]{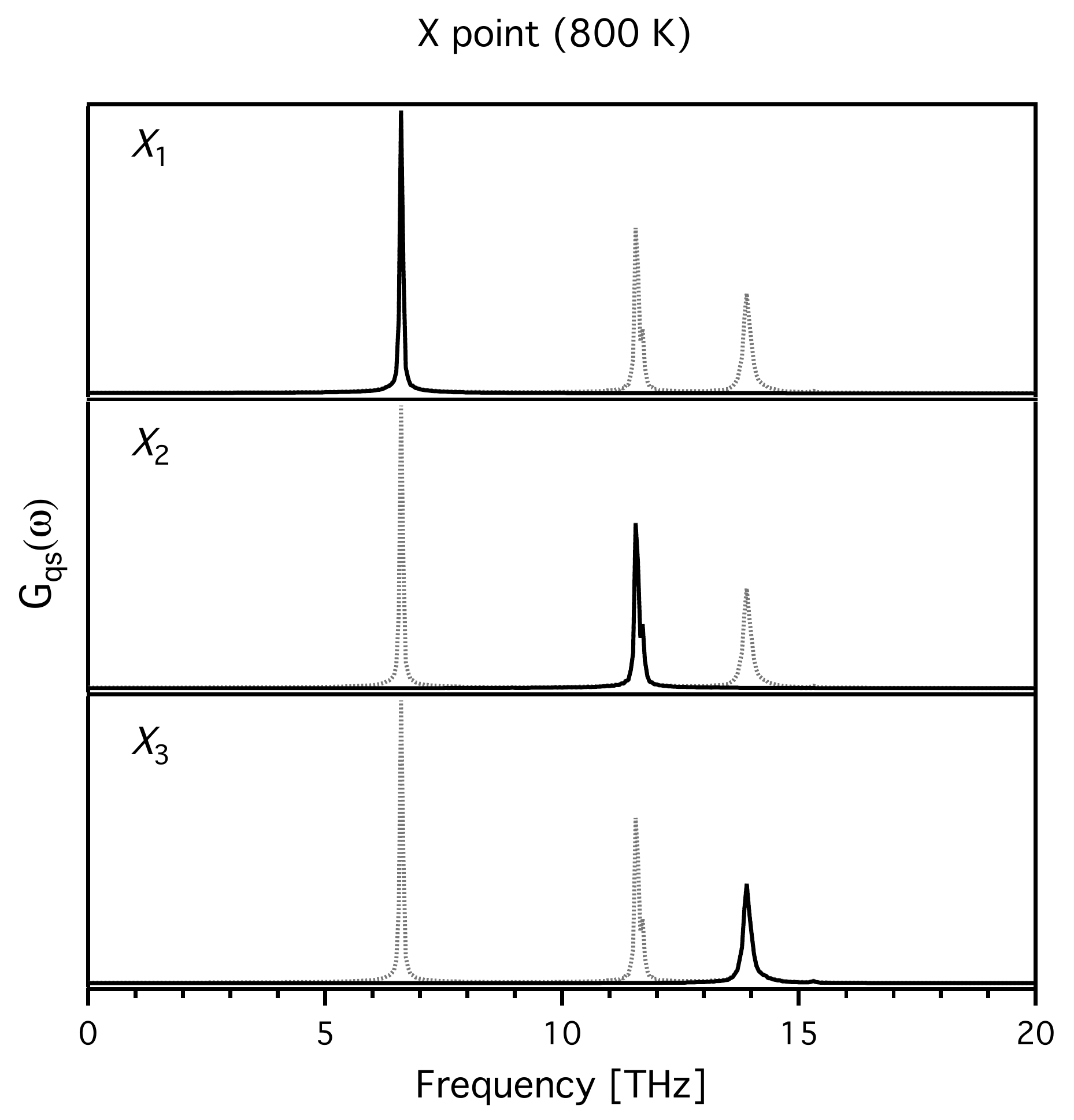}
        \end{subfigure}
        \begin{subfigure}[h]{0.35\textwidth}
                \includegraphics[width=\textwidth]{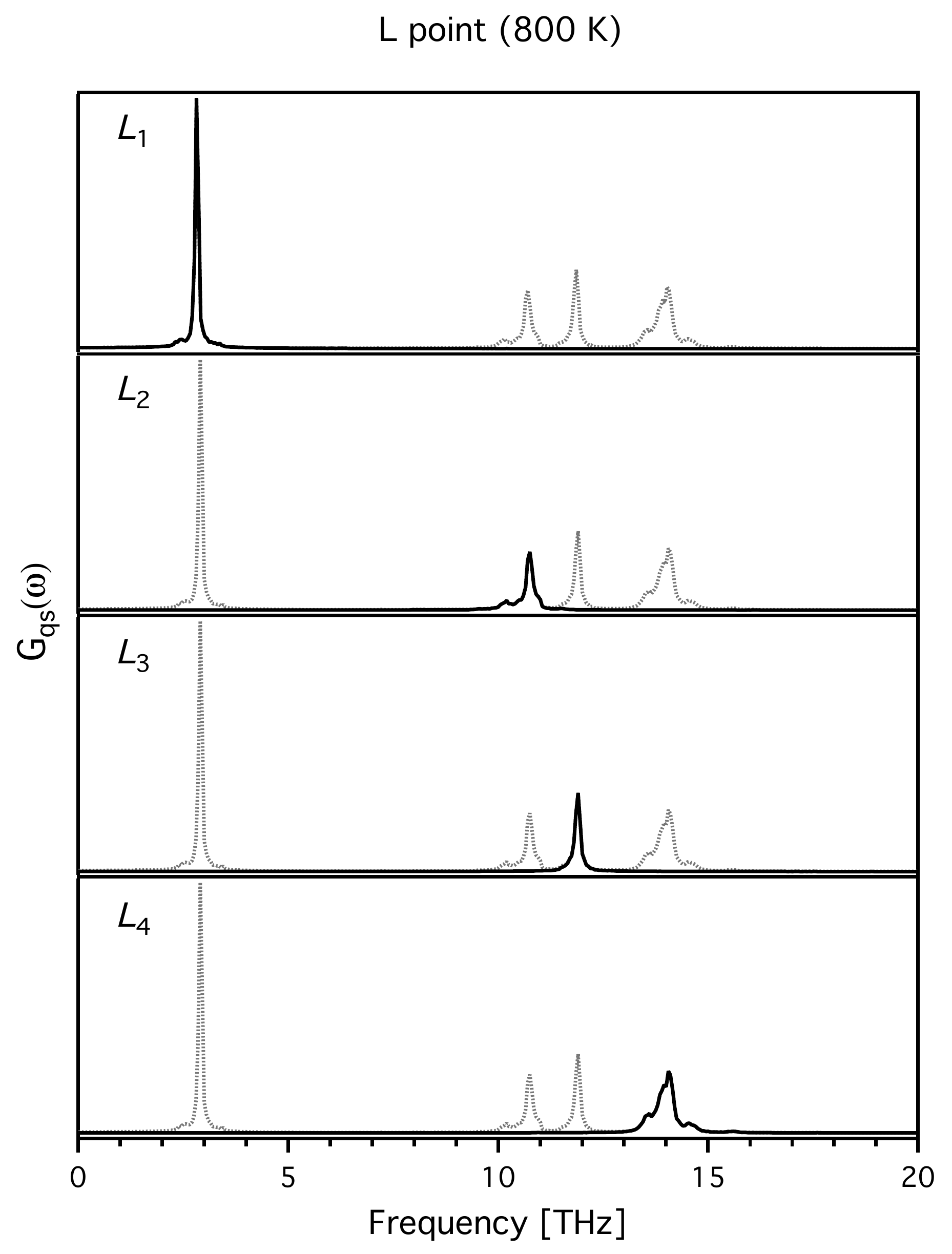}
        \end{subfigure}
        \hspace{2cm}
        \begin{subfigure}[h]{0.35\textwidth}
                \includegraphics[width=\textwidth]{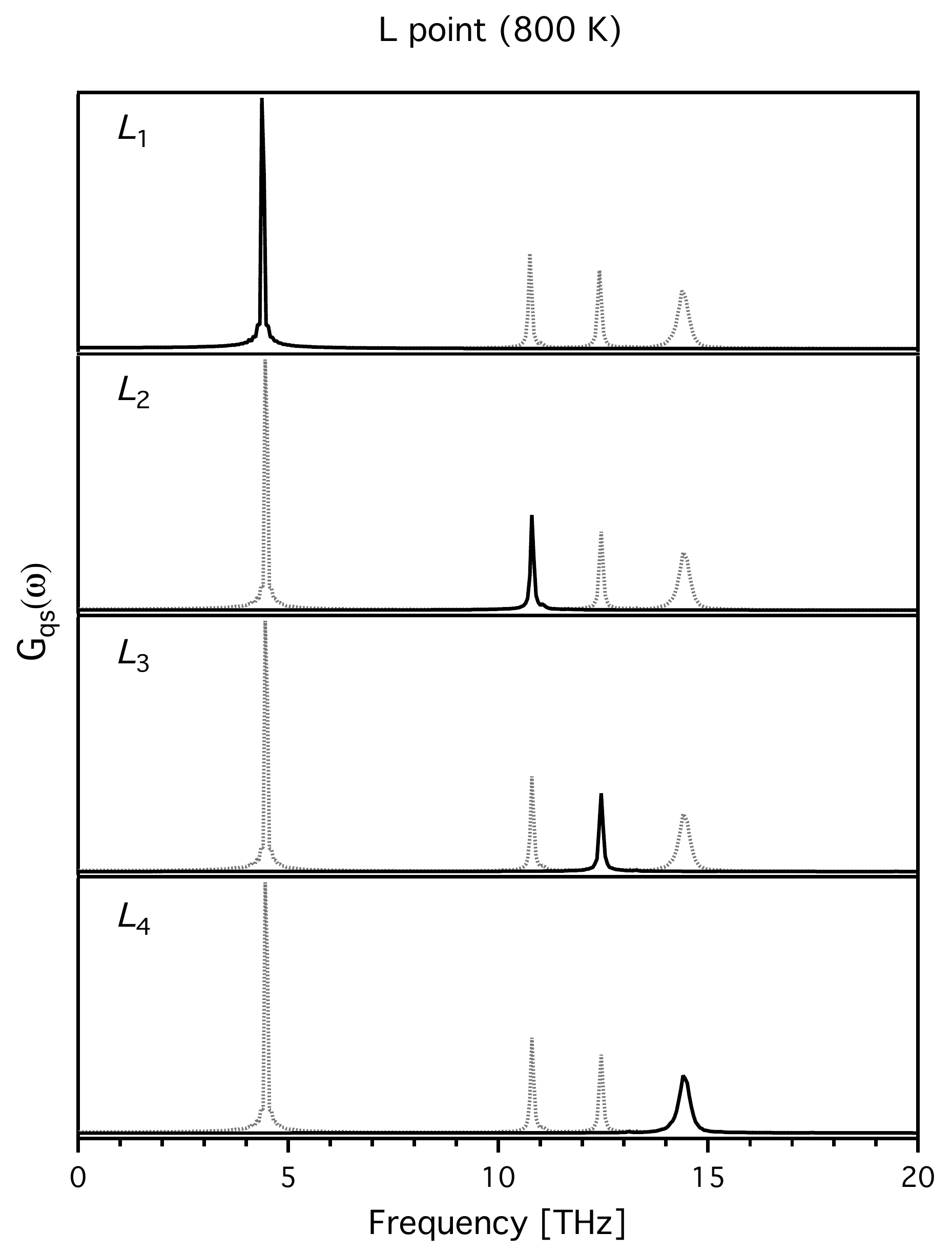}
        \end{subfigure}
        \caption{Wave-vector-projected power spectra $G_{\bf q}(\omega)$ (grey curve) and phonon-mode-projected power spectra $G_{{\bf q}s}(\omega)$ (black curve) of silicon at $\Gamma$, $X$ and $L$. These power spectra have been calculated using the FT method from MD simulations at 800 K computed employing (a) first-principles and (b) the Tersoff empirical potential approaches. Phonon modes are numbered from the lowest to highest phonon frequency.}
        \label{fig:phonon_modes}
\end{figure*}

Fig. \ref{fig:phonon_modes} shows the power spectra $G_{\bf q}(\omega)$ and $G_{{\bf q}s}(\omega)$ at $\Gamma$, $X$ and $L$ points calculated using first-principles and the Tersoff empirical potential. We observe similar results in both approaches. In both cases the projection on the distinguishable phonon modes show a single peak with a Lorenzian function shape (Eq. (\ref{eq:lorentzian_fit})). We observe that each $G_{{\bf q}s}(\omega)$ matches with one of the peaks in the corresponding $G_{\bf q}(\omega)$. This shows that the phonon mode decomposition of the MD trajectories has been performed successfully.

\subsection{Frequency shifts}\label{sec:frequency_shifts}

Here frequency shift refers to the change of the phonon frequency with respect to temperature. At constant pressure condition, the frequency shift is composed of two main contributions: the intrinsic lattice anharmonicity $\Delta \omega^\text{A}_{{\bf q}s}$ and the crystal thermal expansion $\Delta \omega^\text{E}_{{\bf q}s}$. We approximate the total frequency shift $\Delta \omega_{{\bf q}s}$ as 
\begin{linenomath}
	\begin{equation}\label{eq:total_frequency_shift}
		\Delta \omega_{{\bf q}s} \approx \Delta \omega^\text{A}_{{\bf q}s} + \Delta \omega^\text{E}_{{\bf q}s},
	\end{equation}
\end{linenomath}
where $\Delta \omega^\text{A}_{{\bf q}s}$ and $\Delta \omega^\text{E}_{{\bf q}s}$ are computed separately. 
$\Delta \omega^\text{A}_{{\bf q}s}$ is calculated from MD simulations at constant volume conditions and it is measured with respect to the corresponding harmonic frequency obtained by the lattice dynamics calculation.
$\Delta \omega^\text{E}_{{\bf q}s}$ is calculated based on QHA by the following equation: 
 
\begin{linenomath}	
	\begin{equation}\label{eq:frequency_shifts_qha}
		\Delta \omega^\text{E}_{{\bf q}s}(T) =  \omega_{{\bf q}s}[V(T)] - \omega_{{\bf q}s}(V^0),
	\end{equation}
\end{linenomath}
where $V(T)$ is the equilibrium cell volume at the given temperature $T$ and zero pressure, which is obtained by the QHA calculation using \textsc{phonopy-qha}, and $V^0$ the reference cell volume at 0 K and zero pressure.
 
\begin{figure}[h]
        \centering
        \begin{subfigure}[h]{0.23\textwidth}
                \caption{First-principles}
                \includegraphics[width=\textwidth]{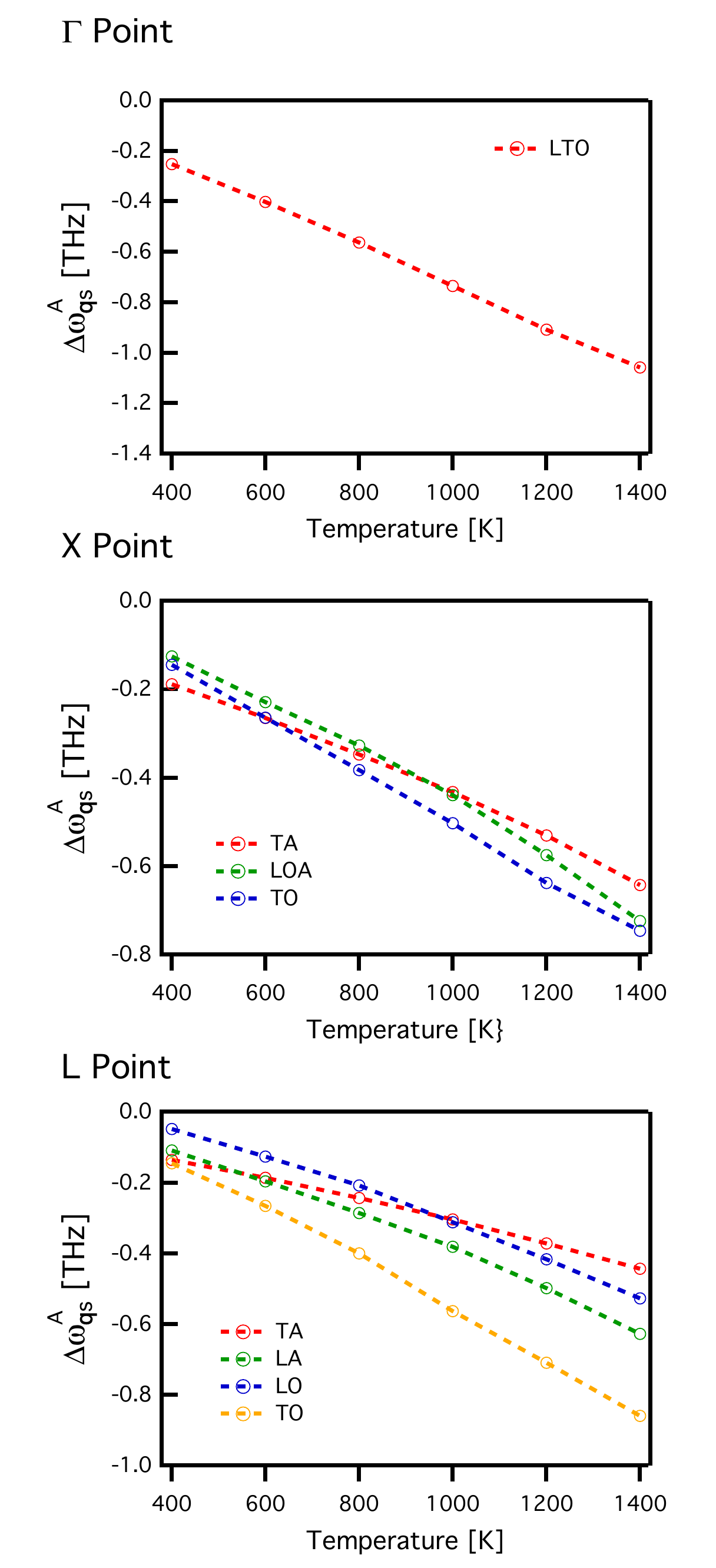}
        \end{subfigure}
        \begin{subfigure}[h]{0.23\textwidth}
                \caption{Empirical potential}
                \includegraphics[width=\textwidth]{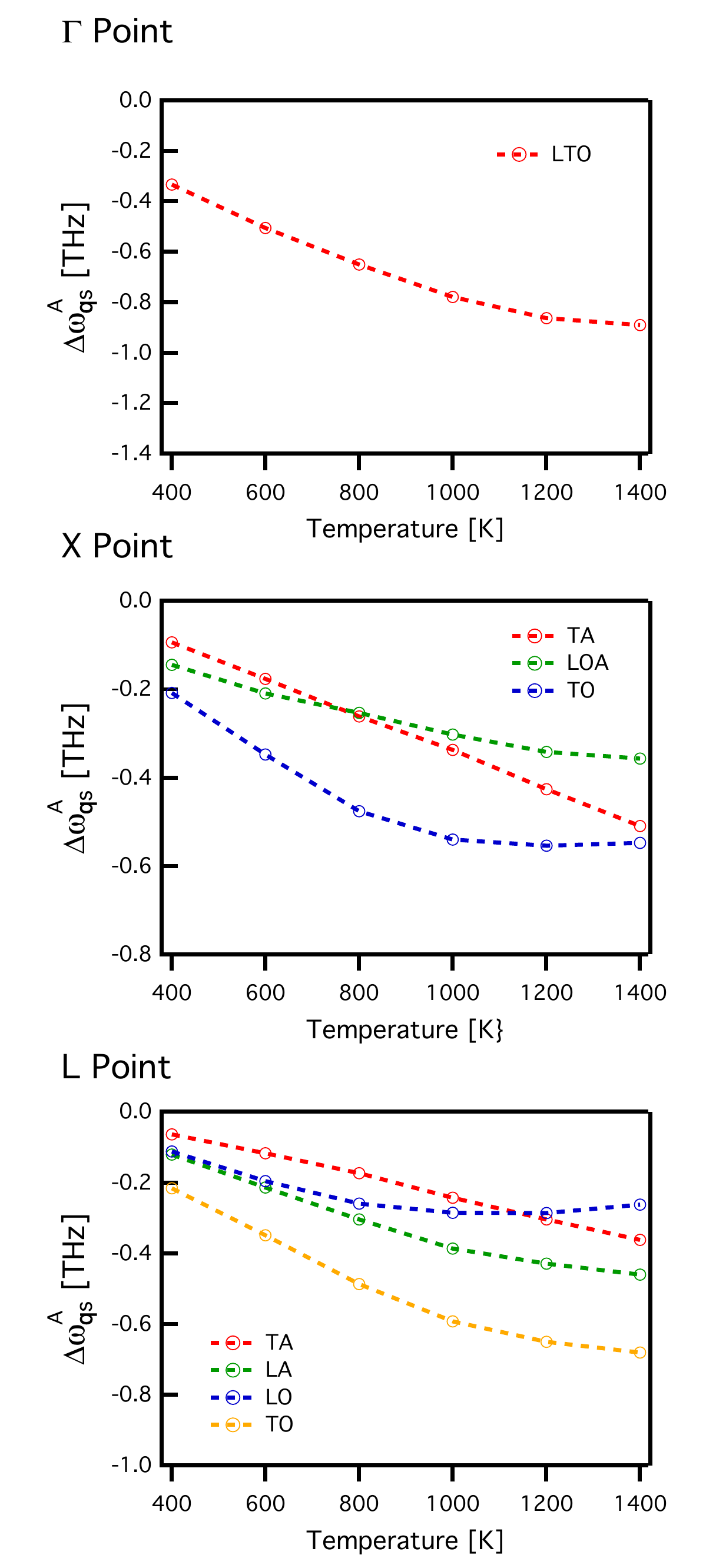}
        \end{subfigure}
        \caption{Intrinsic lattice anharmonicity contribution to the frequency shift of silicon as a function of temperature. These results have been calculated using (a) first-principles and (b) the Tersoff empirical potential approaches. The FT method was used to calculate the frequency shifts depicted as open circles. Dashed lines are guides to the eye.}
        \label{fig:si_shift_vasp}

\end{figure}

In Fig. \ref{fig:si_shift_vasp} we show $\Delta \omega^\text{A}_{{\bf q}s}$ at $\Gamma$, $X$ and $L$ points as a function of temperature obtained using first-principles and the Tersoff empirical potential. 
We observe a qualitative agreement between both approaches. The tendency of $\Delta \omega^\text{A}_{{\bf q}s}$ with the temperature is similar and the orders of the different branches mostly agree. At these three wave vectors, the frequency shifts show a linear tendency with negative slope. This is an expected behavior of the frequency shift when the anharmonicity is weak, i.e., $\Delta \omega_{{\bf q}s} \ll \omega_{{\bf q}s}$. However we observe an unexpected curvature of the frequency shift tendency of some optical modes at high temperatures in the results obtained using the Tersoff empirical potential. This behavior is not observed in the first-principles calculations. 

\begin{figure}[h]
        \centering
        \begin{subfigure}[h]{0.23\textwidth}
                \caption{First-principles}
                \includegraphics[width=\textwidth]{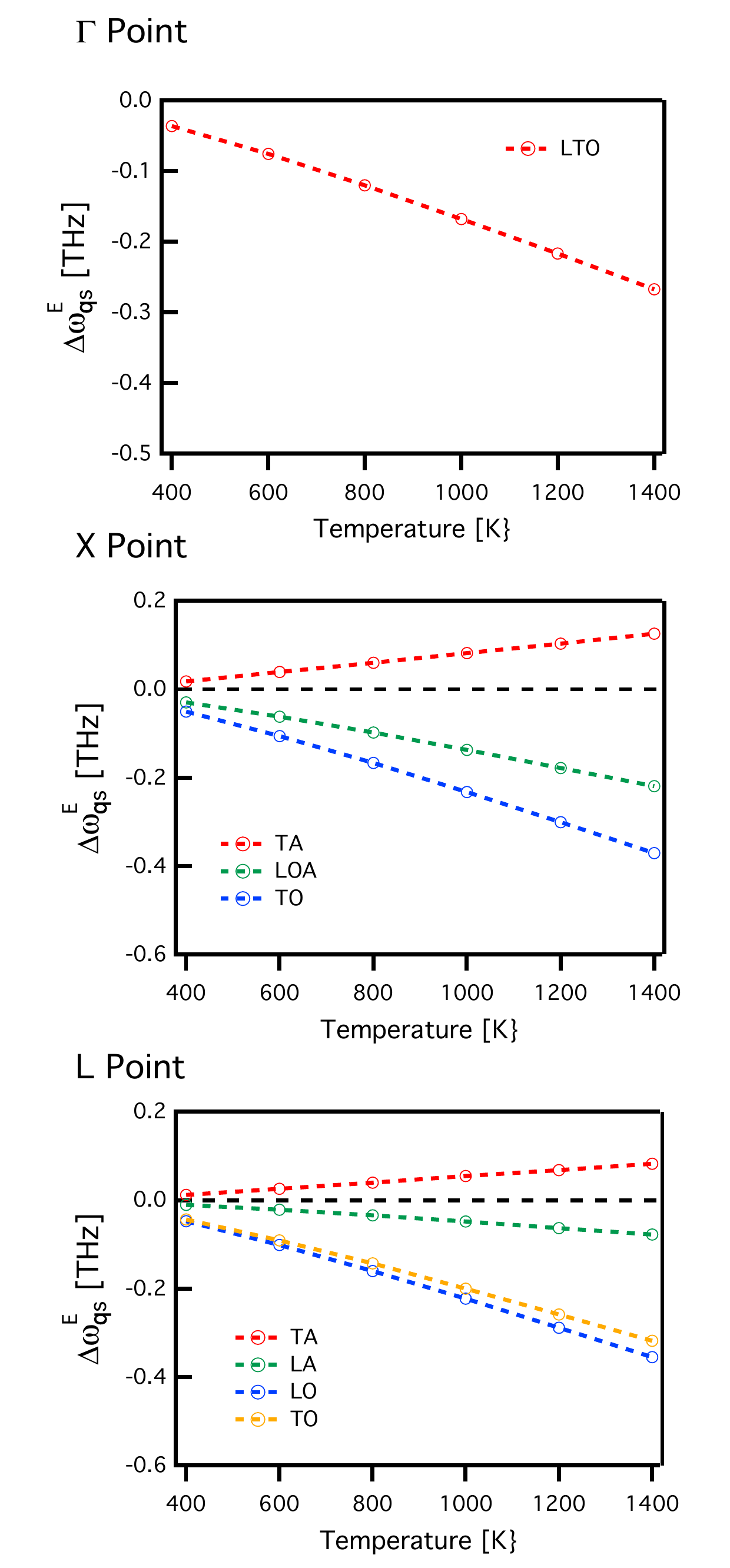}
        \end{subfigure}
        \begin{subfigure}[h]{0.23\textwidth}
                \caption{Empirical potential}
                \includegraphics[width=\textwidth]{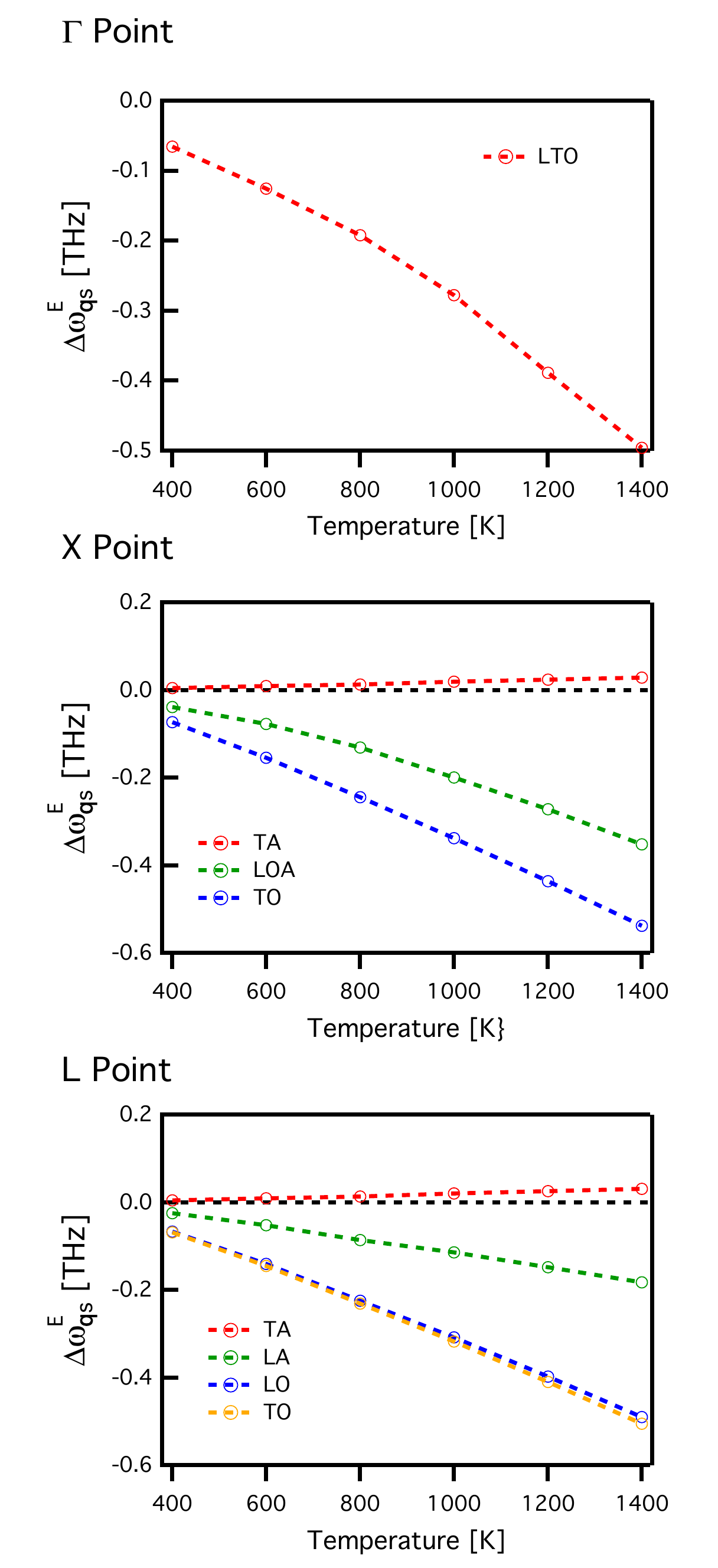}
        \end{subfigure}
        \caption{Thermal expansion contribution to the frequency shifts of silicon calculated using (a) first-principles and (b) the Tersoff empirical potential approaches. MD simulations were computed using the supercell of 64 atoms. Open circles depict the calculations. Dashed lines are guides to the eye.}
        \label{fig:si_volume}
\end{figure}

In Fig. \ref{fig:si_volume} we show $\Delta \omega^\text{E}_{{\bf q}s}$ obtained using first-principles and the Tersoff empirical potential.
In general, we also observe a good qualitative agreement between both approaches. It has to be noted, however, that the results obtained using the Tersoff empirical potential show systematically lower $\Delta \omega^\text{E}_{{\bf q}s}$ than those obtained by first-principles . This is specially noticeable in the LTO branch at $\Gamma$ point and the TA branches at $X$ and $L$ points. 

\begin{figure}[h]
        \centering
        \begin{subfigure}[h]{0.23\textwidth}
                \caption{First-principles}
                \includegraphics[width=\textwidth]{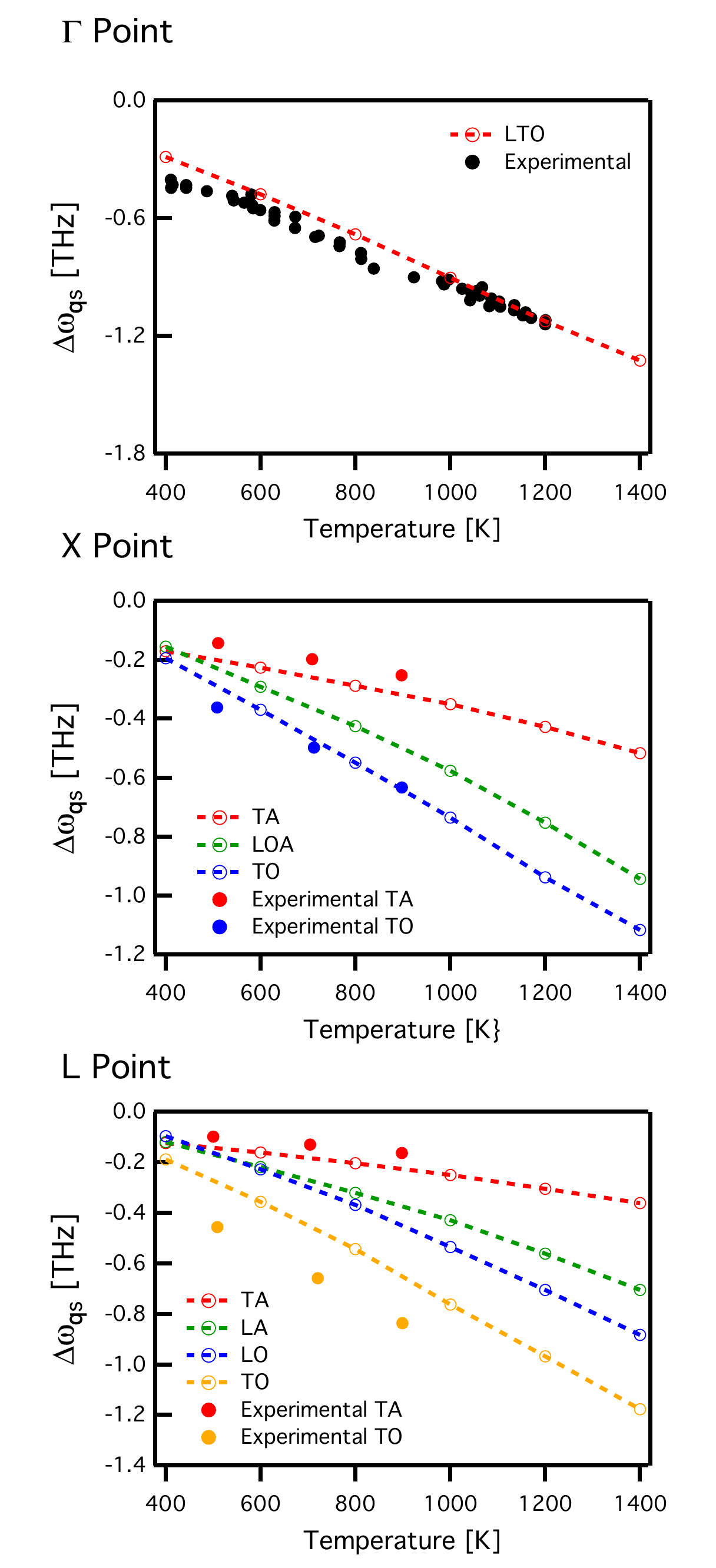}
        \end{subfigure}
        \begin{subfigure}[h]{0.23\textwidth}
                \caption{Empirical potential}
                \includegraphics[width=\textwidth]{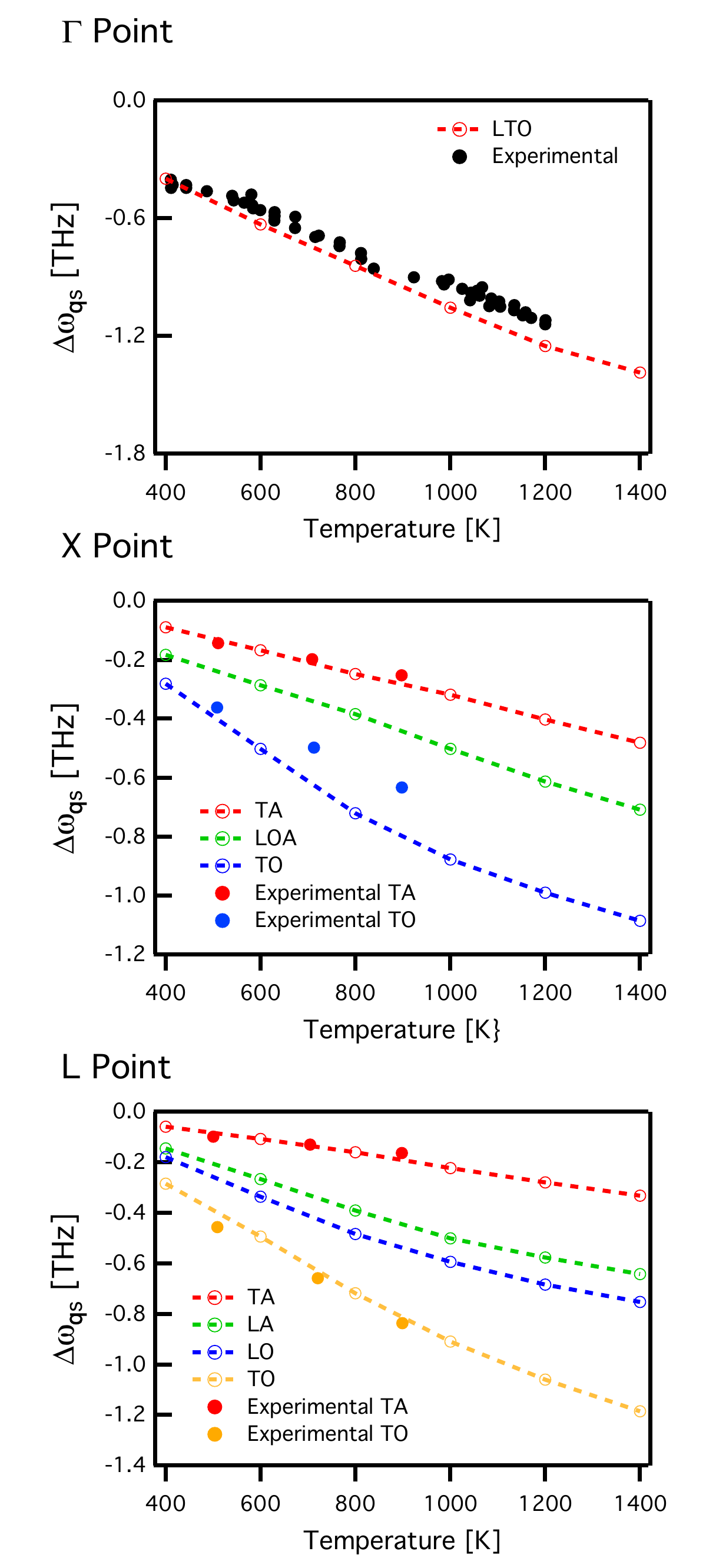}
        \end{subfigure}
        \caption{(a) Total phonon frequency shifts including both of the intrinsic lattice anharmonicity and thermal expansion contributions of silicon obtained using (a) first-principles and (b) the Tersoff empirical potential approaches. MD simulations were computed using the supercell of 64 atoms. Open circles depict the calculations and solid dots are the Raman experimental data from Ref. \cite{Balkanski1983} and ($\Gamma$ point) and Ref. \cite{Tsu1982} ($X$ and $L$ points). Dashed lines are guides to the eye.}
        \label{fig:si_total_shifts}
\end{figure}

In Fig. \ref{fig:si_total_shifts} we show the computed results obtained for the total frequency shift $\Delta \omega_{{\bf q}s}$ compared with experimental results obtained by Raman spectroscopy. The experimental frequency shift is measured with respect to the corresponding estimated harmonic frequency at 0 K as described by Cardona and Ruf\cite{Cardona2001}. Adding the thermal expansion contribution to the frequency shift calculated using the Tersoff empirical potential affects mostly to the acoustic modes. This cancels the curvature observed at high temperatures resulting in a more linear tendency with the temperature. The results obtained from the MD simulations show in general a good agreement with the experimental data.  We obtain an excellent agreement at $\Gamma$ point using both first-principles and the Tersoff empirical potential approaches and we find very similar tendencies with the temperature at $X$ and $L$ points.

\begin{figure}[h]
        \centering
        \begin{subfigure}[h]{0.23\textwidth}
                \caption{64 atoms}
                \includegraphics[width=\textwidth]{si_lammps_volume_64.pdf}
        \end{subfigure}
        \begin{subfigure}[h]{0.23\textwidth}
                \caption{512 atoms}
                \includegraphics[width=\textwidth]{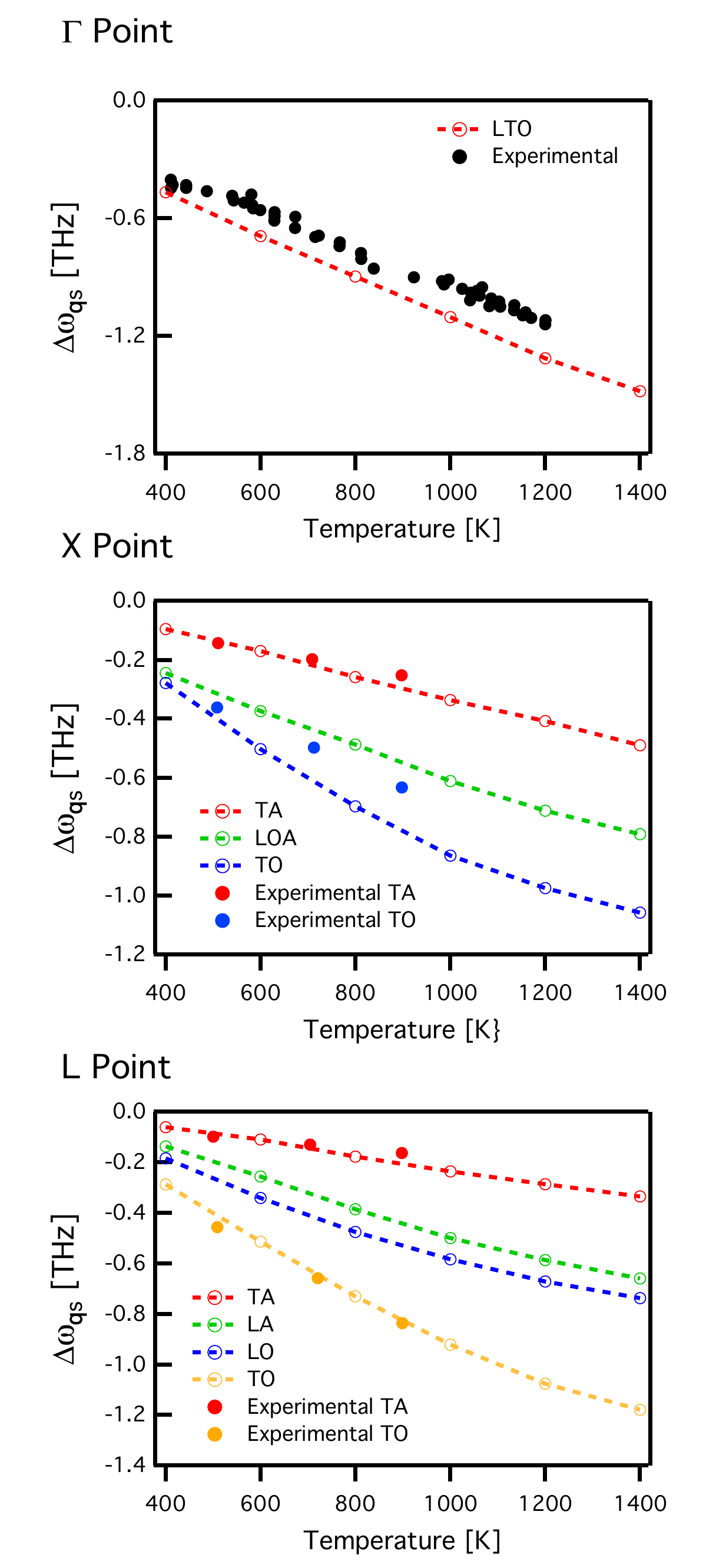}
        \end{subfigure}
        \caption{(a) Total phonon frequency shifts of silicon obtained from the MD simulations using the Tersoff empirical potential and the supercell of (a) 64 atoms and (b) 512 atoms. Open circles depict the calculations and solid dots are the Raman experimental data from Ref. \cite{Balkanski1983} and ($\Gamma$ point) and Ref. \cite{Tsu1982} ($X$ and $L$ points). Dashed lines are guides to the eye.}
        \label{fig:si_total_shifts_large}
\end{figure}

To analyze the supercell size effect on the frequency shifts we calculated the frequency shifts from MD simulations using the Tersoff empirical potential and the supercell of 512 atoms. In Fig. \ref{fig:si_total_shifts_large} we compare the results obtained using the supercells of 64 and 512 atoms. As can be observed, the differences between them are not significant. Therefore we consider that the supercell size effect on the frequency shifts is small and the 64 atoms supercell is sufficient to describe the frequency shift of silicon. This is attributed to short interaction range among atoms in silicon. 

\subsection{Phonon linewidths}\label{sec:linewidths}

\begin{figure}[h]
        \centering
        \begin{subfigure}[h]{0.23\textwidth}
                \caption{First-principles}
                \includegraphics[width=\textwidth]{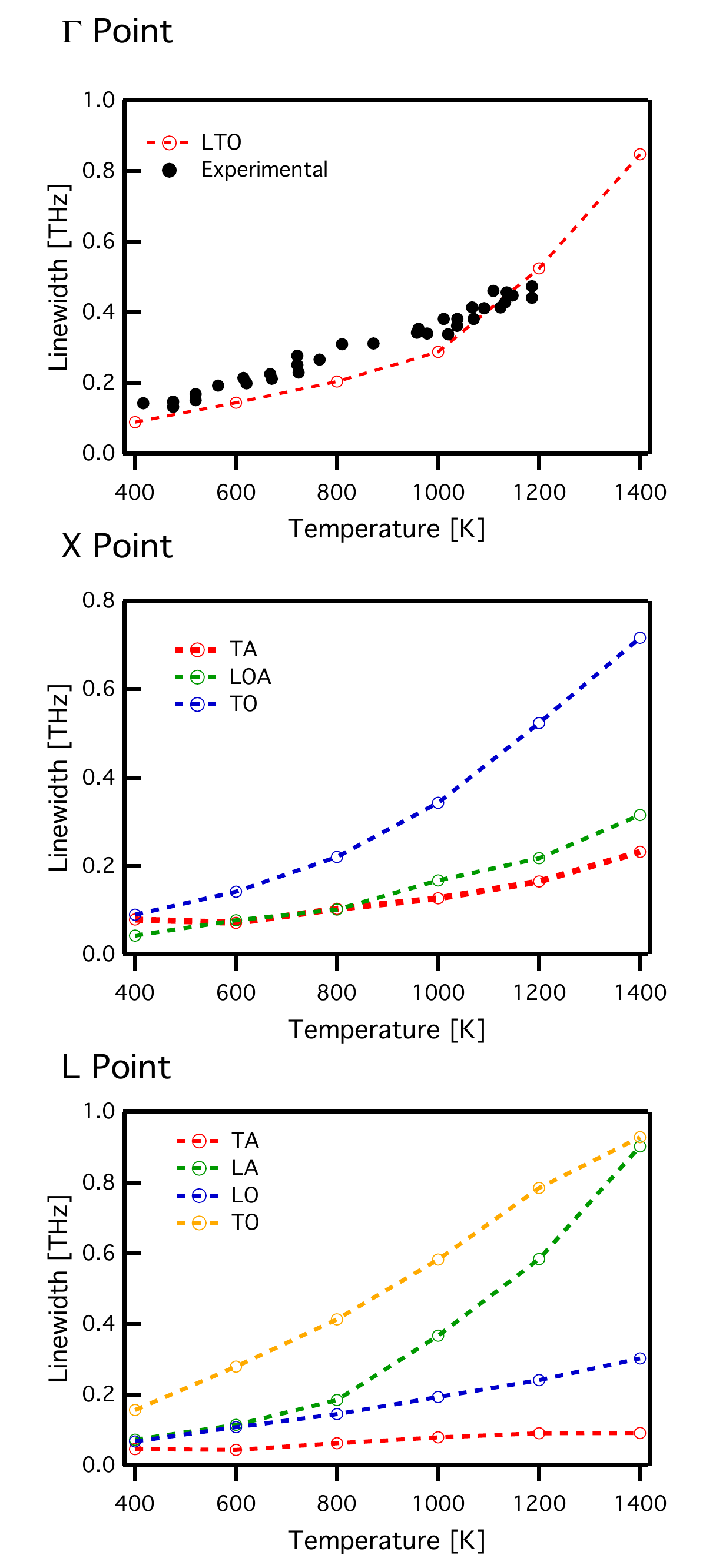}
        \end{subfigure}
        \begin{subfigure}[h]{0.23\textwidth}
                \caption{Empirical potential}
                \includegraphics[width=\textwidth]{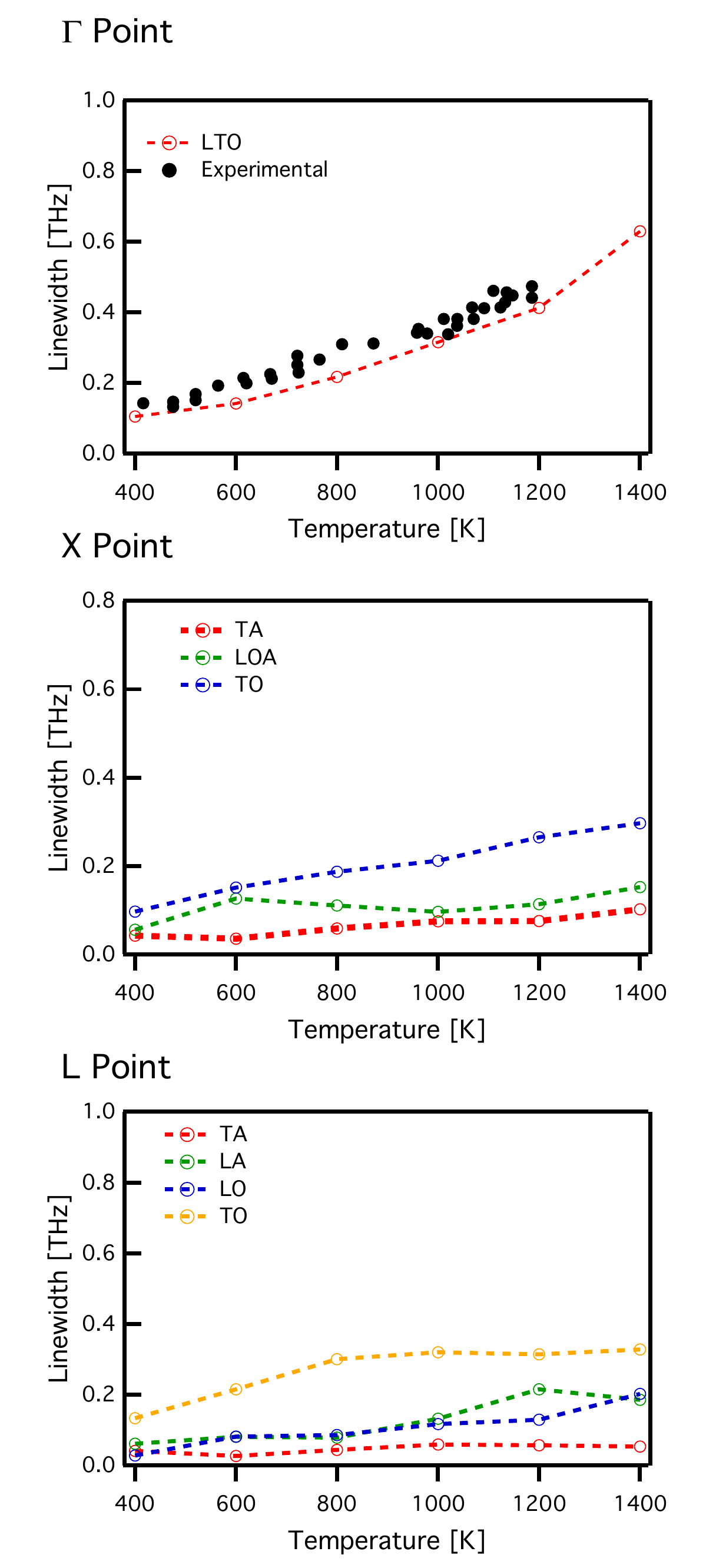}
        \end{subfigure}
        \caption{Linewidths of silicon as a function of temperature obtained from MD simulations using (a) first-principles and (b) the Tersoff empirical potential approaches. MD simulations were computed using the supercell of 64 atoms and the FT method was employed to calculate the power spectra. Open circles depict the calculations and solid dots are the Raman experimental data obtained from Ref. \cite{Balkanski1983} ($\Gamma$ point) and Ref. \cite{Tsu1982} ($X$ and $L$ points). Dashed lines are guides to the eye.}
        \label{fig:si_linewidths_vasp}

\end{figure}

Using MD simulations we computed the phonon linewidths measuring the FWHM of the peak obtained from each $G_{{\bf q}s}(\omega)$ fitted to the Lorentzian function of Eq. (\ref{eq:lorentzian_fit}).
The calculated phonon linewidths are shown in Fig. \ref{fig:si_linewidths_vasp}. In this figure we compare the results obtained using first-principles with the ones obtained using the Tersoff empirical potential. We also include the experimental results obtained by Raman spectroscopy at $\Gamma$ point. As can be seen in Fig. \ref{fig:si_linewidths_vasp} the linewidths calculated using both approaches agree with the experimental result at $\Gamma$ point. However we observe a discrepancy between the results calculated using first-principles and the Tersoff empirical potential approaches at $X$ and $L$ points. The linewidths obtained for the TO and LA branches using first-principles show much larger linewidths at high temperatures than the ones calculated using the Tersoff empirical potential. Unfortunately, we were not able to find experimental results for those wave vectors in literatures.

\begin{figure}[h]
        \centering
        \begin{subfigure}[h]{0.23\textwidth}
                \label{fig:si_shift_lammps_64}
                \caption{64 atoms}
                \includegraphics[width=\textwidth]{si_linewidths_64_lammps.pdf}
        \end{subfigure}
        \begin{subfigure}[h]{0.23\textwidth}
                \caption{512 atoms}
                \includegraphics[width=\textwidth]{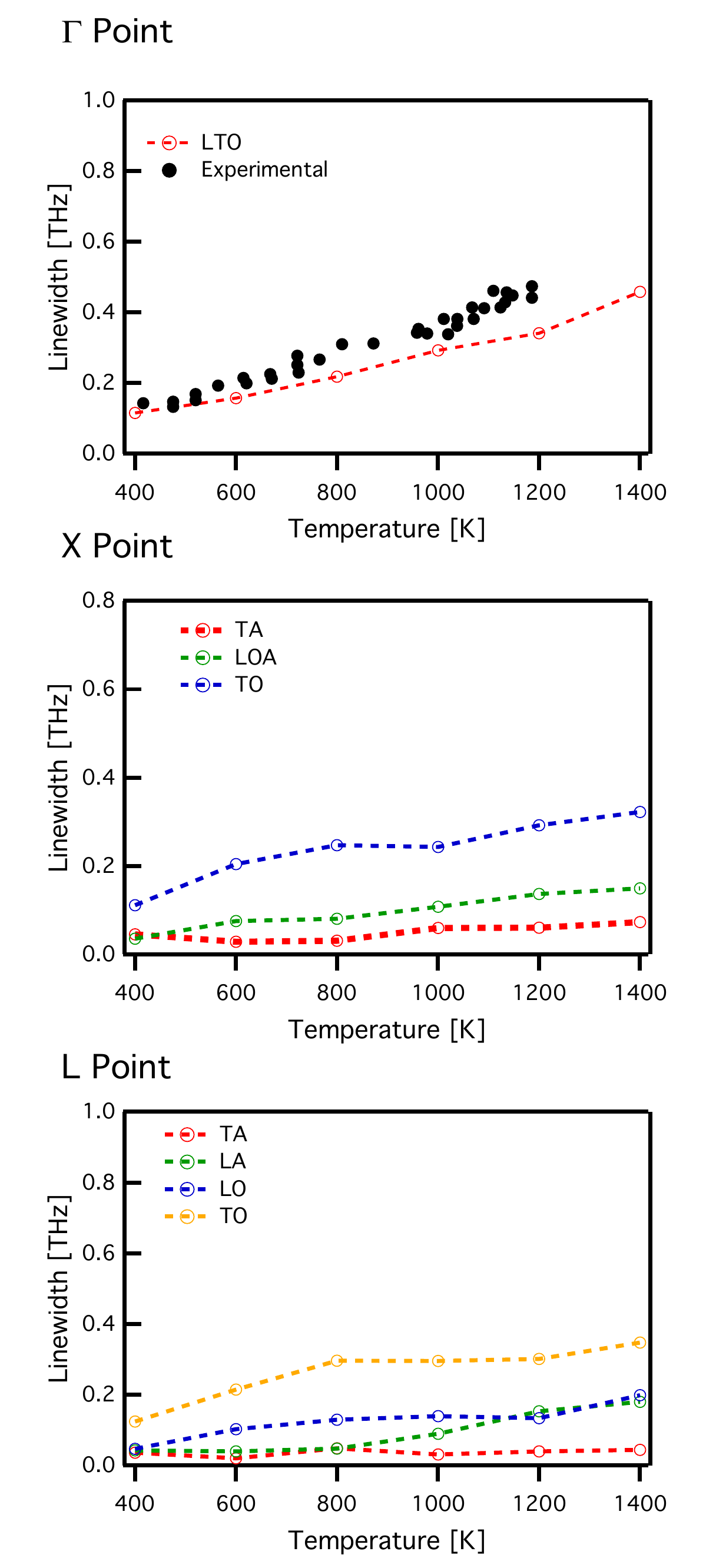}
        \end{subfigure}
        \caption{Phonon linewidths of silicon as a function of temperature. Linewidths have been obtained from MD simulations computed with the Tersoff empirical potential using the supercell of (a) 64 atoms and (b) 512 atoms. Open circles depict the calculations and solid dots are the Raman experimental data obtained from Ref. \cite{Balkanski1983} ($\Gamma$ point) and Ref. \cite{Tsu1982} ($X$ and $L$ points). Dashed lines are guides to the eye}
        \label{fig:si_linewidths_lammps}
\end{figure}

To analyze the effect of the supercell size on the linewidth we employed the Tersoff empirical potential to compute MD simulations using the supercell of 512 atoms. 
In Fig. \ref{fig:si_linewidths_lammps} we compare the phonon linewidths obtained using the supercells of 64 and 512 atoms.  
In this figure we observe a weak dependence of the linewith with the supercell size. Despite small differences observed between the two supercell sizes, we consider that they occur because the width measured from the fitted power spectrum is very sensitive to small changes in its spectral shape. Therefore, we do not expect large differences in the linewidth using larger supercells and we think that the supercell of 64 atoms is sufficient to analyze the linewidths of silicon with our acceptable accuracy.

\section{Conclusions}

We have implemented the phonon-mode-decomposition technique following the scheme proposed by Sun et. al\cite{Sun2014}. 
This methodology provides a systematic procedure to calculate the microscopic phonon properties including anharmonicity to arbitrary order from MD simulations.

To test our implementation we computed the frequency shifts and linewidths of crystalline silicon at $\Gamma$, $X$ and $L$ points using first-principles and the Tersoff empirical potential approaches. 
Using first-principles we obtained a good agreement of the frequency shift and linewidth with the experiments of Raman spectroscopy.
In spite of the limitations of empirical potentials, the frequency shifts calculated using the Tersoff empirical potential showed a good agreement with the experiments and the first-principles calculations. However, the results obtained for the linewidths showed some discrepancies with respect to the first-principles calculations at the $X$ and $L$ points. 

This study was also a good test of the treatment of anharmonicity on the Tersoff empirical potential. We believe that this methodology and our software are useful not only to study anharmonic materials but also to develop and clarify new empirical potentials that considers the anharmonicity. 

At the present time, first-principles MD simulations are still computationally expensive. However, due to the increase in power of computers, we expect that these calculations will be easily accessible in the near future.

\section{Acknowledgments}
This study was supported by a Grant-in-Aid for Scientific Research on Innovative Areas ``Nano Informatics'' (Grant No. 25106005) from the Japan Society for the Promotion of Science (JSPS). We also would like to express our appreciation to Terumasa Tadano for fruitful discussions.

\section{References}
\bibliography{dynaphopy}

\begin{thebibliography}{10}
\expandafter\ifx\csname url\endcsname\relax
  \def\url#1{\texttt{#1}}\fi
\expandafter\ifx\csname urlprefix\endcsname\relax\def\urlprefix{URL }\fi
\expandafter\ifx\csname href\endcsname\relax
  \def\href#1#2{#2} \def\path#1{#1}\fi

\bibitem{dove1993introduction}
M.~T. Dove, Introduction to lattice dynamics, Vol.~4, Cambridge university
  press, 1993.

\bibitem{wallace1998thermodynamics}
D.~C. Wallace, Thermodynamics of crystals, Courier Corporation, 1998.

\bibitem{McGaughey2013}
A.~McGaughey, Predicting phonon properties from equilibrium {Molecular
  Dynamics} simulations, Annu. Rev. Heat Transf. (2013) 49--87.

\bibitem{Sun2014}
T.~Sun, D.~Zhang, R.~Wentzcovitch, Dynamic stabilization of cubic {CaSiO$_3$}
  perovskite at high temperatures and pressures from ab initio molecular
  dynamics, Phys. Rev. B Condens. Matter Mater. Phys. 89 (2014)
  094109/1--094109/9.

\bibitem{Tersoff1988}
J.~Tersoff, Empirical interatomic potential for silicon with improved elastic
  properties, Phys. Rev. B 38~(14) (1988) 9902.

\bibitem{Balkanski1983}
M.~Balkanski, R.~Wallis, E.~Haro, Anharmonic effects in light scattering due to
  optical phonons in silicon, Phys. Rev. B 28~(4) (1983) 1928--1934.

\bibitem{Tsu1982}
R.~Tsu, J.~G. Hernandez, Temperature dependence of silicon raman lines, Appl.
  Phys. Lett. 41~(11) (1982) 1016--1018.

\bibitem{Zhang2014}
D.~B. Zhang, T.~Sun, R.~M. Wentzcovitch, Phonon quasiparticles and anharmonic
  free energy in complex systems, Phys. Rev. Lett. 112~(5) (2014) 1--5.

\bibitem{Sun2010a}
T.~Sun, X.~Shen, P.~B. Allen, Phonon quasiparticles and anharmonic perturbation
  theory tested by molecular dynamics on a model system, Phys. Rev. B -
  Condens. Matter Mater. Phys. 82~(22) (2010) 1--10.

\bibitem{Lee1993}
C.~Lee, D.~Vanderbilt, K.~Laasonen, R.~Car, M.~Parrinello, Ab initio studies on
  the structural and dynamical properties of ice, Phys. Rev. B 47~(9) (1993)
  4863--4872.

\bibitem{vasp1996}
G.~Kresse, J.~Furthmuller, Efficient iterative schemes for \textit{ab initio}
  total-energy calculations using a plane-wave basis set, Phys. Rev. B 54
  (1996) 11169--11186.

\bibitem{vasp1999}
G.~Kresse, D.~Joubert, From ultrasoft pseudopotentials to the projector
  augmented-wave method, Phys. Rev. B 59 (1999) 1758--1775.

\bibitem{Plimpton1995a}
S.~Plimpton, Fast parallel algorithms for short-range {Molecular Dynamics}, J.
  Comput. Phys. 117~(1) (1995) 1--19.

\bibitem{phonopy}
A.~Togo, F.~Oba, I.~Tanaka, First-principles calculations of the ferroelastic
  transition between rutile-type and {CaCl$_2$}-type {SiO$_2$} at high
  pressures, Phys. rev. B 78 (2008) 134106.

\bibitem{Allen2015}
P.~B. Allen, Anharmonic phonon quasiparticle theory of zero-point and thermal
  shifts in insulators: Heat capacity, bulk modulus, and thermal expansion,
  Phys. Rev. B - Condens. Matter Mater. Phys. 92~(6) (2015) 1--11.

\bibitem{Monkhorst1976}
H.~J. Monkhorst, J.~D. Pack, Special points for brillouin-zone integrations,
  Phys. Rev. B 13 (1976) 5188--5192.

\bibitem{PBESol2008}
J.~P. Perdew, A.~Ruzsinszky, G.~I. Csonka, O.~A. Vydrov, G.~E. Scuseria, L.~A.
  Constantin, X.~Zhou, K.~Burke, Restoring the density-gradient expansion for
  exchange in solids and surfaces, Phys. Rev. Lett. 100 (2008) 136406.

\bibitem{Hoover1695}
W.~G. Hoover, Canonical dynamics: Equilibrium phase-space distributions, Phys.
  Rev. A 31 (1985) 1695--1697.

\bibitem{Cardona2001}
M.~Cardona, T.~Ruf, Phonon self-energies in semiconductors: Anharmonic and
  isotopic contributions, Solid State Commun. 117~(3) (2001) 201--212.

\bibitem{VanDerWalt2011}
S.~Van Der~Walt, S.~C. Colbert, G.~Varoquaux, The numpy array: A structure for
  efficient numerical computation, Comput. Sci. Eng. 13~(2) (2011) 22--30.

\bibitem{Frigo1999a}
M.~Frigo, A fast fourier transform compiler, ACM SIGPLAN Notices 34~(5) (1999)
  169--180.

\bibitem{press2007numerical}
W.~H. Press, Numerical recipes 3rd edition: The art of scientific computing,
  Cambridge university press, 2007, Ch.~13.

\bibitem{Benesty2008}
J.~Benesty, J.~Chen, Y.~A. Huang, Springer Handbook of Speech Processing,
  Springer Berlin Heidelberg, 2008, Ch.~7.

\end{thebibliography}
\appendix
\section{The discrete Fourier transform} \label{sec:dft}
In the MD calculations the velocity is obtained at a constant time interval $\Delta$. 
We use the discrete Fourier transform (FT) to compute the power spectrum from the discretized data. The discrete FT is defined for a finite number of samples $h_n$ as

\begin{linenomath}
	\begin{equation} \label{eq:discrete_fourier_transform}
		H_j\equiv \sum\limits^{N_\mathrm{s}-1}_{n=0}h_n e^{i \omega n / N_\mathrm{s}},
	\end{equation}
\end{linenomath}
where $N_\mathrm{s}$ is the number of samples.
In this definition, the inverse discrete Fourier transform that recovers $h_n$ from $H_j$ is
\begin{linenomath}
	\begin{equation} \label{eq:iDFT}
		h_n =  \frac{1}{2 \pi N_\mathrm{s}}\sum_{j=0}^{N_\mathrm{s}-1}H_j e^{-i \omega n / N_\mathrm{s}}.
	\end{equation}
\end{linenomath}
The Fourier transform $H(\omega_j)$ is estimated from the discrete FT as 
\begin{linenomath}
	\begin{equation} \label{eq:DFT_to_FT}
		H(\omega_j) = \int_{-\infty}^{\infty}h(t)e^{i \omega_j t} dt \approx \sum_{n=0}^{N_\mathrm{s}-1}h_n e^{i \omega_j t_n} \Delta  = \Delta H_j,
	\end{equation}
\end{linenomath}
where
\begin{linenomath}
	\begin{equation} \label{eq:DFT_time}
		t_n \equiv n\Delta.
	\end{equation}
\end{linenomath}
$H(\omega_j)$ is obtained at discrete frequencies $\omega_j$ given by 
\begin{linenomath}
	\begin{equation} \label{eq:f_j}
		\omega_j = \frac{2\pi j}{N_\mathrm{s}\Delta},\qquad j=-\frac{N_\mathrm{s}}2,...,\frac{N_\mathrm{s}}2.
	\end{equation}
\end{linenomath}
The extreme values of $j$ in Eq. (\ref{eq:f_j}) correspond to the lower and upper limits at which the discrete FT is defined.
The corresponding frequency at these limits is known as the Nyquist frequency $\omega_\mathrm{N}$ that is 
\begin{linenomath}
	\begin{equation} \label{eq:nyquist_frequency}
		\omega_\mathrm{N} = \frac{\pi}{\Delta}.
	\end{equation}
\end{linenomath}
The power spectrum computed from the discrete FT is obtained at the same resolution and defined within the same frequency range than the discrete FT.
Therefore, the resolution at which the power spectrum is obtained is given by eq. (\ref{eq:f_j}) and corresponds to
\begin{linenomath}
	\begin{equation} \label{eq:resolution}
		\omega_\mathrm{R} = \frac{2\pi}{N_{\mathrm{s}}\Delta}.
	\end{equation}
\end{linenomath}
In contrast to $\omega_\mathrm{N}$, $\omega_\mathrm{R}$ depends on the number of samples $N_\mathrm{s}$. At constant $\Delta$, increasing $N_\mathrm{s}$ improves the resolution of $H(\omega_j)$. 

To make the best use of the MD data, we carefully chose the value of $\omega_\mathrm{N}$ and $\omega_\mathrm{R}$ to obtain power spectra defined within the frequency range that we are interested in, and with enough resolution to resolve the peaks that correspond to the quasiparticle phonon modes.
In this study we set $\omega_\mathrm{R}$ to 0.05 THz, which is smaller than the width of the peaks that we want to measure and $\omega_\mathrm{N}$ is determined by the time step used to compute the MD simulations.
According to Eq. (\ref{eq:resolution}), the number of samples $N_\mathrm{s}'$ needed to calculate the power spectrum with this resolution is
\begin{linenomath}
	\begin{equation} \label{eq:f_n2}
		N_\mathrm{s}' = \frac{2\pi}{0.05 \Delta}.
	\end{equation}
\end{linenomath}
Because the number of time steps calculated from MD simulations is larger than $N_\mathrm{s}'$, we divide the full trajectory in small segments of length $N_\mathrm{s}'$. Then we calculate the power spectrum of each segment and average them to obtain the final power spectrum. By this procedure we improve the precision of the measurement of the power spectrum from the MD data.

We use the fast Fourier transform (FFT) algorithm to calculate the  discrete FT. This algorithm allow us to reduce the computation time significantly compared with the direct method shown in Eq. (\ref{eq:powerspectrum}). There exists several implementations of this algorithm. In \textsc{DynaPhoPy} we use numpy\cite{VanDerWalt2011} and FFTW\cite{Frigo1999a} implementations.

\section{The maximum entropy method} \label{sec:mem}
An alternative approach to estimate the power spectrum is the maximum entropy (ME) method\cite{press2007numerical}. This method allow us to obtain a smoother power spectrum estimation with lower computational cost than using the discrete FT. However, according to our tests, it requires a larger number of samples than the discrete FT to obtain a good estimation. In this section we summarize the algorithm employed in our code.

In the ME method we assume that the atoms trajectory follows an autoregressive model. Using this model, the power spectrum is calculated as
\begin{linenomath}
	\begin{equation} \label{eq:powermem}
		P(\omega) = \frac{{ \langle x^2\rangle}\Delta}{{{{\left| {1 - \sum\limits_{j = 1}^m {{a_j}{e^{i \omega \Delta j}}} } \right|}^2}}},
	\end{equation}
\end{linenomath}
where ${\{a_j\}}$ form a set of $m$ coefficients that defines the model, $\langle x^2 \rangle$ is the mean square discrepancy between the model and the data and $\Delta$ is the sampling interval. The number of coefficients is an input parameter in this method and it has to be carefully chosen. These coefficients are calculated from a linear prediction that is a mathematical operation where the subsequent values $h_n$ are predicted by a linear function of the previous ones using a set of coefficients $a_j$. The linear predictor is defined as

\begin{linenomath}
	\begin{equation}\label{eq:model_forward}	
		{h_n} = \sum\limits_{j = 1}^k {{a_j}{h_{n - j}}}+f_{k,n},
	\end{equation}
\end{linenomath}
where $f_{k,n}$ is the discrepancy between the original data and the prediction. 
This is called the forward linear prediction. This operation can also be done in the opposite direction, predicting the former values as a function of the following ones obtaining the backward prediction:
\begin{linenomath}
	\begin{equation}\label{eq:model_backward}	
		{h_{n-k}} = \sum\limits_{j = 1}^k {{c_j}{h_{n - j + 1}}}+b_{k,n},
	\end{equation}
\end{linenomath}
where $b_{k,n}$ is the discrepancy in the backward linear prediction and $c_j$ are the backward prediction coefficients. These coefficients are related to $a_j$ by

\begin{linenomath}
	\begin{equation}\label{eq:coefficients_relation}	
		c_j = a_{k-j+1}.
	\end{equation}
\end{linenomath}

The aim of the ME method is to obtain the best set of coefficients that minimizes the total discrepancy. To archive this we used the Burg's method\cite{press2007numerical}, where the sum of the forwards and backward discrepancy is minimized using the Levinson-Durbin algorithm\cite{Benesty2008}. In this algorithm the coefficients $a_j$ are determined through a recursive procedure in which the coefficients are related each other by parameters $\mu_k$ that are optimized at each iteration. The parameters $\mu_k$ are known as reflection coefficients and are obtained by

\begin{linenomath}	
	\begin{equation}\label{eq:reflection_coefficients}	
		\mu_k = \frac {2{\bf b}^T_k {\bf f}_k}{{\bf f}^T_k {\bf f}_k + {\bf b} ^T_k {\bf b}_k}.
	\end{equation}
\end{linenomath}
${\bf f}_k$ and ${\bf b}_k$ are known as the discrepancy vectors and are defined as

\begin{linenomath}	
	\begin{equation}\label{eq:coef_a}
		{\bf f}_k = \left( {\begin{array}{*{20}{c}}
		f_{k,2}\\
		f_{k,3}\\
		f_{k,4}\\
 		\vdots \\
		f_{k,N}
		\end{array}} \right) \;\;\;\;\;\;\;\;\;
		{\bf b}_k = \left( {\begin{array}{*{20}{c}}
		b_{k,1}\\
		b_{k,2}\\
		b_{k,3}\\
 		\vdots \\
		b_{k,N-1}
		\end{array}} \right),
	\end{equation}
\end{linenomath}
where
\begin{linenomath}
	\begin{equation}\label{eq:model_forward_0}	
		f_{0,n} = b_{0,n} = {h_n}.
	\end{equation}
\end{linenomath}
The Burg's method algorithm proceeds as follows:
\begin{enumerate}

\item Initialization
\begin {algorithmic}
\State $a_0 \gets {2{\bf b}^T_0 {\bf f}_0} / ({{\bf f}^T_0 {\bf f}_0 + {\bf b} ^T_0 {\bf b}_0})$
\State $k \gets 1$
\end{algorithmic}

\item Calculation of the reflection coefficient
\begin {algorithmic}
\State $\mu_k \gets {2{\bf b}^T_k {\bf f}_k} / ({{\bf f}^T_k {\bf f}_k + {\bf b} ^T_k {\bf b}_k})$
\end{algorithmic}

	\item Update of coefficients $a_j$

\begin {algorithmic}
\State $a_{j+1}\gets0$
\For {$j \gets 0, k-1$}
    \State $a_j\gets a_j+\mu_k a_{k-1-j}$
\EndFor
\end{algorithmic}

\item Increment of $k$
\begin {algorithmic}
\If {$k \leq m$}
        \State $k\gets k+1$
		\State {Back to step 2}
\EndIf
\end{algorithmic}

\item Finish

\end{enumerate}
As can be observed, at each iteration $k$ a new set of coefficients $a_j$ is calculated until $k=m$ is reached. Finally, the mean square discrepancy is calculated from the reflection coefficients $\mu_k$ as
\begin{linenomath}	
	\begin{equation}\label{eq:mean_discrepancy}	
		\langle x^2 \rangle= \frac{1}N_\mathrm{s}\sum\limits_{n = 0}^{N_\mathrm{s}-1} {h^2_n} - \sum\limits_{k = 1}^{m}\mu_k.
	\end{equation}
\end{linenomath}

\section{Supercell size effect on first-principles MD calculations} \label{sec:mem}
\begin{figure*}[t]
        \centering
        \begin{subfigure}[h]{0.33\textwidth}
                \caption{Frequency shift}
                \includegraphics[width=\textwidth]{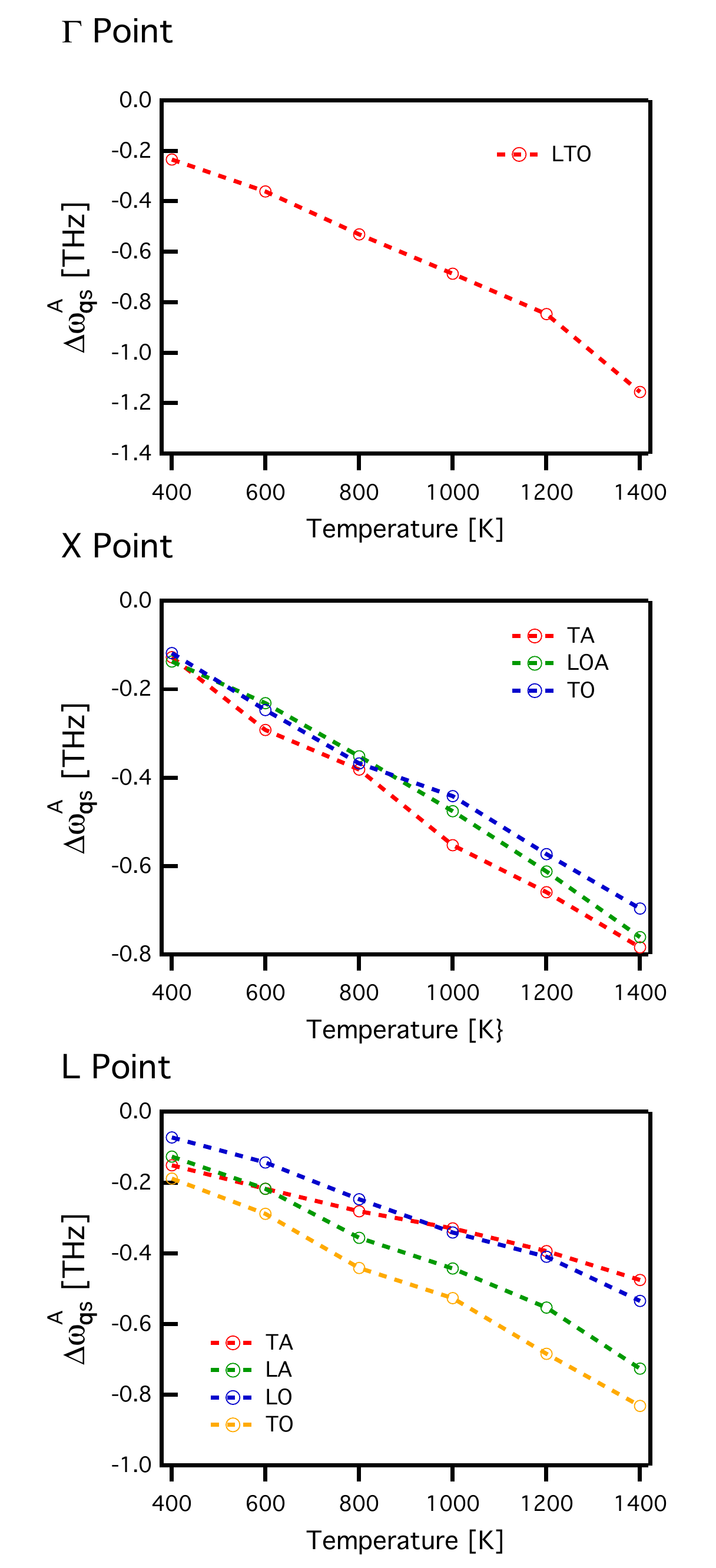}
        \end{subfigure}
        \begin{subfigure}[h]{0.33\textwidth}
                \caption{Total shift}
                \includegraphics[width=\textwidth]{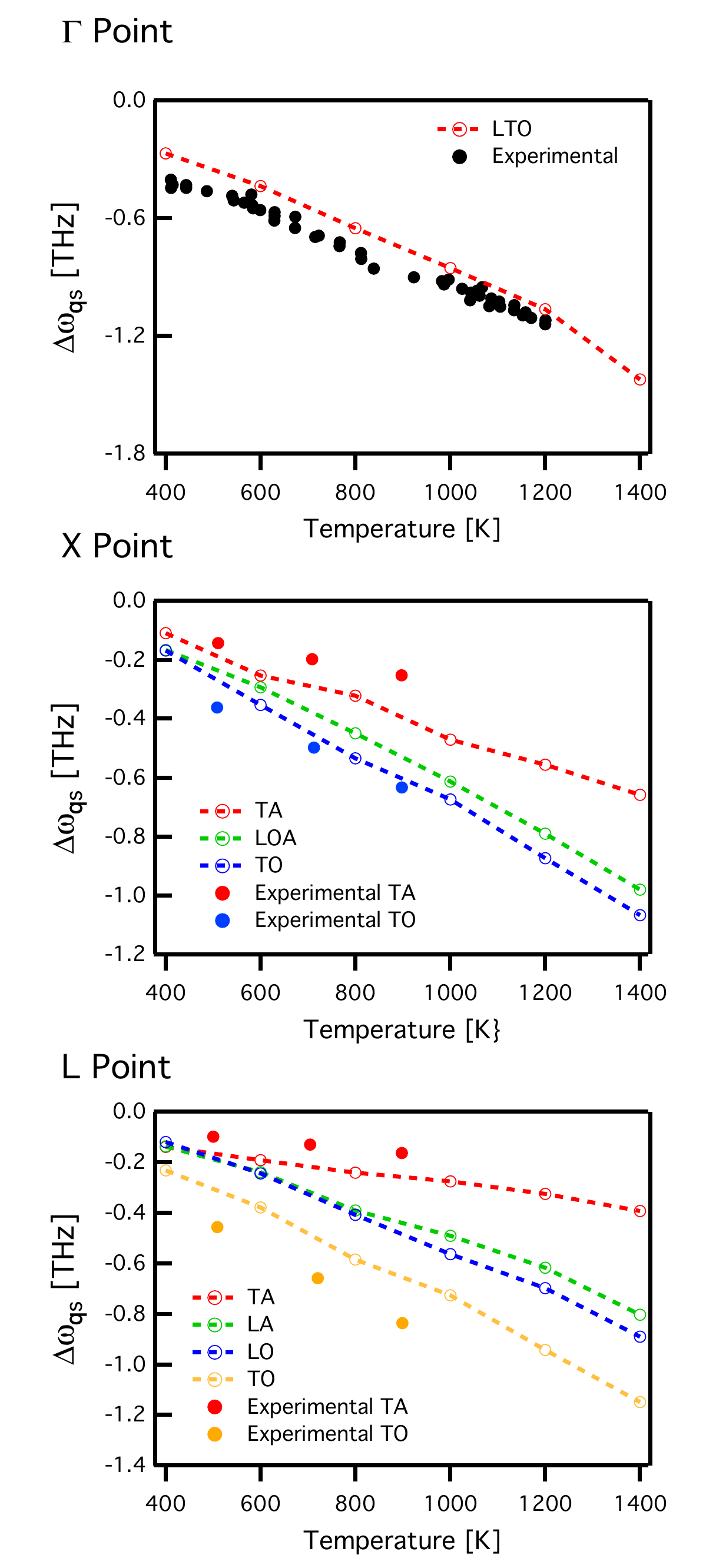}
        \end{subfigure}
        \begin{subfigure}[h]{0.33\textwidth}
                \caption{Linewidth}
                \includegraphics[width=\textwidth]{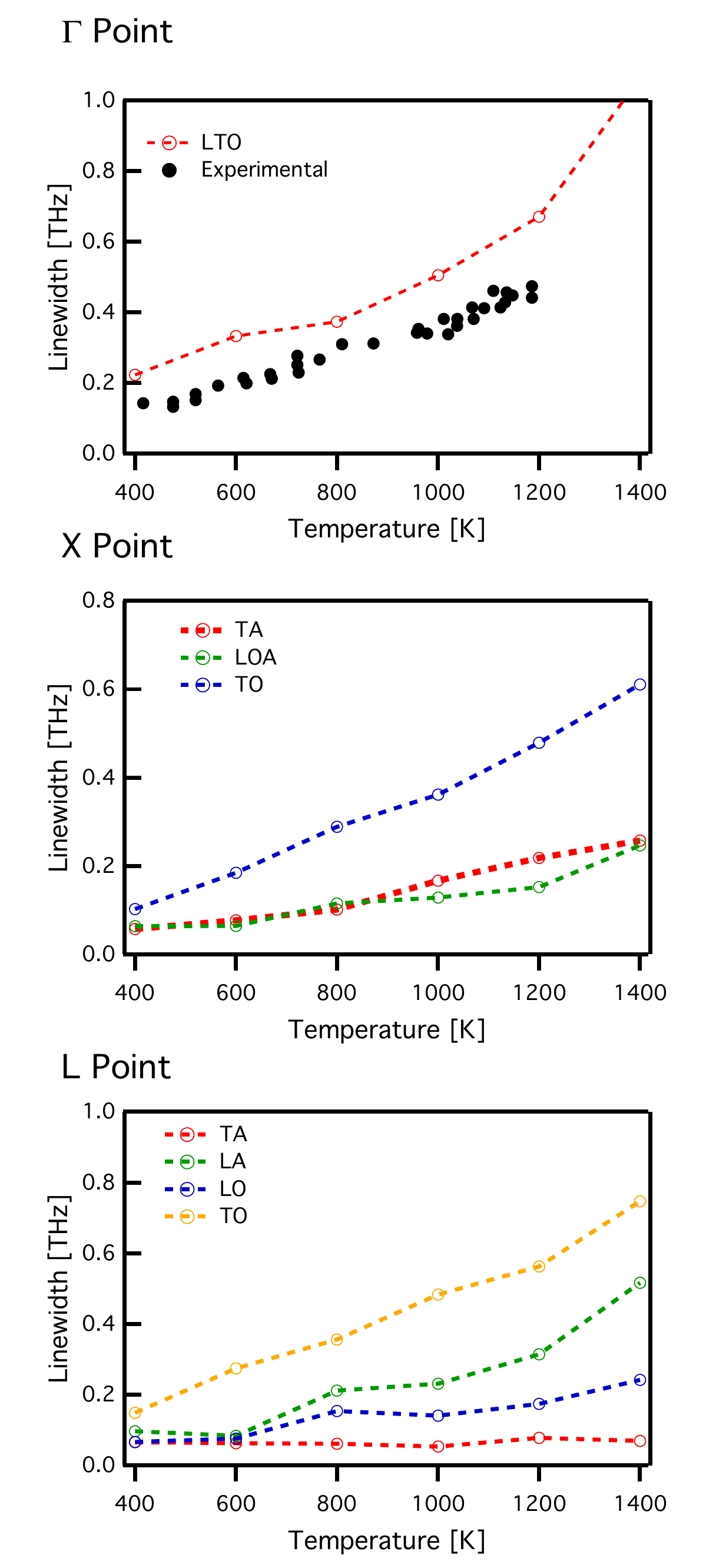}
        \end{subfigure}
        \caption{Intrinsic lattice anharmonicity contribution to the frequency shifts (a), total frequency shifts (b) and linewidths (c) of silicon at $\Gamma$, $X$ and $L$ points obtained from MD simulations computed using first-principles approach and the supercell of 512 atoms. The FT method was used to extract the quasiparticle phonon properties. Dashed lines are guides to the eye.}
        \label{fig:si_512}
\end{figure*}

We have also analyzed the effect of the supercell size on the quasiparticle phonon properties using first-principles calculations by calculating MD simulations using the supercell of 512 atoms. Unfortunately due to the high computational demand of these calculations we are not able to show results obtained with the same precision as the ones presented in the main text. For this reason, especially in the case of the linewidths, these results should be taken with caution and be used as qualitative references. In Fig. \ref{fig:si_512} we show the frequency shifts and linewidths of Si calculated from MD using the supercell of 512 atoms. The power spectrum was calculated using the FT method employing 30000 time steps. In this figure we observe similar results for the frequency shifts than the ones obtained using the 64 atoms supercell. Consequently, the total shifts computed including the thermal expansion contribution are also in agreement with the experiments as can be observed in Fig. \ref{fig:si_512}c. This allow us to conclude that, as observed in the analysis shown in Sec. \ref{sec:frequency_shifts} using the Tersoff empirical potential, the supercell of 64 atoms is sufficient to compute the frequency shifts of Si.

For linewidths, however, we observe that in the result at $\Gamma$ point obtained using the supercell of 512 atoms, although shows the same tendency with the temperature, is slightly overestimated with respect to the experiment. At the other analyzed high symmetry points, $X$ and $L$, we observe also small differences with respect to the calculations using the supercell of 64 atoms but with similar tendencies with the temperature. It is interesting to notice, however, that the linewidths obtained using the supercell of 512 atoms show a more linear tendency with the temperature than the ones obtained using the supercell of 64 atoms. This behavior is in agreement with the general linear tendency that linewidths usually present with the temperature in experiments. This suggest that the curvature that the phonon linewidths show in Fig. \ref{fig:si_linewidths_vasp}, especially at high temperatures,  may be an effect of to the limited supercell size. This effect can also be observed in the linewidth at $\Gamma$ point obtained using the Tersoff empirical potential approach shown in Fig. \ref{fig:si_linewidths_lammps}.

\end{document}